\documentclass[a4paper]{article}

\usepackage{amssymb}
\usepackage{amsmath}
\usepackage{amsthm}
\usepackage{verbatim}

\usepackage{hhline}

\usepackage{array}
\usepackage{multirow}
\usepackage{endnotes}

\usepackage{epsf}

\usepackage{rotating}

\input xy
\xyoption{all}
\CompileMatrices

\newtheorem{theorem}{Theorem}[section]
\newtheorem{lemma}[theorem]{Lemma}

\newtheorem{corollary}[theorem]{Corollary}

\theoremstyle{definition}
\newtheorem{definition}[theorem]{Definition}

\theoremstyle{definition}

\theoremstyle{remark}
\newtheorem{remark}[theorem]{Remark}

\theoremstyle{remark}
\newtheorem{example}[theorem]{Example}

\newcommand{\g}{\mathfrak{g}}

\newcommand{\ssl}{\operatorname{sl}}

\newcommand{\id}{{1 \mskip -5mu {\rm I}}}

\newcommand{\vac}{|0\rangle}

\newcommand{\Span}{\mbox{span}}

\newcommand{\Ker}{\mbox{Ker}}

\renewcommand{\footnote}{\endnote}

\newcommand{\T}{\mathcal{T}}
\newcommand{\M}{\mathcal{M}}

\newcommand{\A}{\mathcal{A}}
\newcommand{\II}{\mathcal{I}}
\newcommand{\B}{\mathcal{B}}
\newcommand{\tB}{\tilde{\mathcal{B}}}

\newcommand{\tU}{\tilde{\mathcal{U}}}

\renewcommand{\P}{L}
\newcommand{\N}{N}

\newcommand{\skewl}{\mbox{sl}}
\newcommand{\skewn}{\mbox{sn}}
\newcommand{\lw}{\mbox{W}^{l}}
\newcommand{\rw}{\mbox{W}^{r}}
\newcommand{\qa}{\mbox{Q}}
\renewcommand{\j}{\mbox{J}}

\newcommand{\I}{\mathcal{I}}

\newcommand{\cw}{w}

\newcommand{\bi}{\bar{\text{\itshape{\i}}}}

\newcommand{\bC}{\mathbb C}

     \newcommand{\bZ}{\mathbb Z} 

\newcommand{\R}{\mathbb R}      \newcommand{\Z}{\mathbb Z}
\newcommand{\C}{\mathbb C}
 
\begin{document}

%%%%%%%%%%% TITOLO %%%%%%%%%%%%%%%%%%%%%%%%%%%%

\par\vskip 1cm\vskip 2em

\begin{center}

{\Huge  Freely generated vertex algebras and non-linear Lie conformal algebras}

\par

\vskip 2.5em \lineskip .5em

{\large
\begin{tabular}[t]{c}
$\mbox{ Alberto De Sole}^{1}
\phantom{mmm}\mbox{Victor G. Kac}^{2}$
\end{tabular}
\par
}
\bigskip

{\small
\begin{tabular}[t]{ll}
{\bf 1} & Department of Mathematics, Harvard University \\
& 1 Oxford st, Cambridge, MA 02138, USA \\
& E--mail: {\tt desole@alum.mit.edu}\\
{\bf 2} & Department of Mathematics, MIT \\
& 77 Massachusetts Av, Cambridge, MA 02139, USA \\
& E--mail: {\tt kac@math.mit.edu}
\end{tabular}
}
\bigskip

\end{center}

%%%%%%%%% ABSTRACT %%%%%%%%%%%%%%%%%%%%%%%%%%%%

\vskip 1 em
\centerline{\bf Abstract} 
\smallskip
{\small 
\noindent We introduce the notion of a non--linear Lie conformal superalgebra and prove 
a PBW theorem for its universal enveloping vertex algebra.
We also show that conversely any graded freely generated vertex algebra 
is the universal enveloping algebra of a non--linear Lie conformal superalgebra.
This correspondence will be applied in the subsequent work to the problem of classification
of finitely generated simple graded vertex algebras.
}

\bigskip

%%%%%%%% SECTION 1: INTRODUCTION %%%%%%%%%%%%%%%%%%%%%%%%%%%%%

\section{Introduction}
\setcounter{equation}{0}

After the work of Zamolodchikov \cite{zam} it has become clear that 
the chiral (= vertex) algebra
of a conformal field theory gives rise to a Lie algebra with non--linearities in commutation relations.
This and the subsequent works in the area clearly demonstrated that the absence of non--linearities,
like in the Virasoro algebra, the current algebras and their super analogues, 
is an exception, rather then a rule.

This is closely related to the fact that in the singular parts of the operator product expansions
of generating fields, as a rule, not only linear combinations of these fields and their derivatives
occur, but also their normally ordered products.
In fact, the absence of terms with normally ordered products is equivalent to the absence
of non--linearities in the commutation relations.

The latter case is encoded in the notion of a Lie conformal superalgebra \cite{kac1},
and a complete classification of finite simple Lie conformal superalgebras was given in 
\cite{fattori} (completing thereby a sequence of works that began with \cite{rs}
and continued in the conformal algebra framework 
in \cite{kac2}, \cite{dandrea1} and \cite{fattori}).

A complete classification in the general non--linear case is much harder and is still far away.
In the present paper we lay rigorous grounds to the problem by introducing the notion
of a non--linear conformal superalgebra.

In order to explain the idea, let us define this notion in the quantum mechanical setting.
Let $\g$ be a vector space, endowed with a gradation
$\g=\oplus_j\g_j$, by a discrete subsemigroup of the additive semigroup 
of positive real numbers, and with a linear map
$[\ ,\ ]:\ \g\wedge\g\rightarrow\T(\g)$, where $\T(\g)$ is the tensor algebra over $\g$.
We write $\Delta(a)=j$ if $a\in\g_j$, and we extend the gradation to $\T(\g)$ by additivity,
i.e. $\Delta(1)=0,\ \Delta(A\otimes B)=\Delta(A)+\Delta(B)$, for $A,B\in\T(\g)$.
We denote 
$$
\M_j(\g)\ =\ \Span
\bigg\{
 A\otimes (b\otimes c-c\otimes b-[b,c]) \otimes D
\left|\begin{array}{c}
 b,c\in \g,A,D\in\T(\g)\\
 \Delta(A\otimes b\otimes c\otimes D))\leq j
 \end{array}\right\}
$$
and we let $\M(\g)=\cup_j\M_j(\g)$. Note that $\M(\g)$ is the two sided ideal of the tensor algebra
generated by elements $a\otimes b-b\otimes a-[a,b]$, where $a,b\in\g$.
The map $[\ ,\ ]$ is called a {\itshape non--linear Lie algebra} if it satisfies the following
two properties ($a,b,c\in\g$):
\begin{eqnarray*}
\vphantom{\Big(}
\text{(grading condition)} && \Delta([a,b])<\Delta(a)+\Delta(b)\ , \\
\vphantom{\Big(}
\text{(Jacobi identity)} && [a,[b,c]]-[b,[a,c]]-[[a,b],c]\in\M_{\Delta}(\g)\ ,\\
\vphantom{\Big(}
&& \qquad\text{ where } \Delta<\Delta(a)+\Delta(b)+\Delta(c)\ .
\end{eqnarray*}
Let $U(\g)=\T(\g)/\M(\g)$. It is not difficult to show by the usual method (see e.g. \cite{jacobson})
that the PBW theorem holds for the associative algebra $U(\g)$, i.e. the images of all
ordered monomials in an ordered basis of $\g$ form a basis of $U(\g)$.

The purpose of the present paper is to introduce the notion of a non--linear Lie conformal 
(super)algebra $R$, which encodes the singular part of general OPE of fields,
in order to construct the corresponding universal enveloping vertex algebra $U(R)$, 
and to establish for $U(R)$ a PBW theorem.

The main difficulty here, as compared to the Lie (super)algebra case, comes from the fact
that $U(R)=\T(R)/\M(R)$, where $\M(R)$ is not a two--sided ideal of the associative algebra $\T(R)$.
This makes the proof of the PBW theorem much more difficult.

Of course, in the "linear" case these results are well known \cite{kac1}, \cite{gms}, \cite{bakalov}.

A simpler result of the paper is a converse theorem, which states that any graded
vertex algebra satisfying the PBW theorem is actually the universal enveloping vertex algebra
of a non--linear Lie conformal superalgebra.

The special case when the vertex algebra is freely generated by a Virasoro field
and primary fields of conformal weight 1 and 3/2 was studied in \cite{thesis}.

Here is a brief outline of the contents of the paper.
After reviewing in Section 2 the definition of a Lie conformal algebra
and a vertex algebra, we introduce in Section 3 the definition of a non--linear Lie conformal
algebra $R$. 
We also state the main result of the paper, a PBW theorem for the universal 
enveloping vertex algebra $U(R)$, which is proved in Sections 4 and 5.
In Section 6 we state and prove the converse statement,
namely to every graded freely generated vertex algebra $V$
we associate a non--linear Lie conformal algebra $R$ such that $V\simeq U(R)$.

The results of this paper have important applications to the classification problem
of conformal vertex algebras.
Such applications will be discussed in subsequent work, \cite{non_lin_2}.

% acknowledgments

We would like to thank M. Artin, B. Bakalov, A. D'andrea and P. Etingof
for useful discussions.
This research was conducted by A. De Sole for the Clay Mathematics Institute.
The paper was partially supported by the NSF grant DMS0201017.

%%%%%%%% SECTION 2 %%%%%%%%%%%%%%%%%%%%%%%%%%%%%

\section{Definitions of Lie conformal algebras and vertex algebras}
\setcounter{equation}{0}

% Def of conformal algebras

\begin{definition}
A {\itshape Lie conformal superalgebra} (see \cite{kac1}) is 
a $\Z/2\Z$--graded $\C[T]$-module $R=R_{\bar{0}}\oplus R_{\bar{1}}$ 
endowed with a $\C$-linear map $R\otimes R \rightarrow \bC[\lambda]\otimes R$ 
denoted by $a\otimes b \mapsto [a \ _\lambda \  b]$ 
and called $\lambda$-bracket, satisfying the following axioms
\begin{equation*}
\begin{array}{ll}
\!\!\!\!\!\!\!\!\!
[T a  \ _\lambda \  b]\ =\ -\lambda[a \ _\lambda \  b] 
  \ ,\quad [a \ _\lambda \ T b]\ =\ (\lambda+T)[a \ _\lambda \ b]
&\ \ \ \text{(sesquilinearity)} \\
\!\!\!\!\!\!\!\!\!
\vphantom{\Bigg(}
{[b  \ _\lambda \  a]}\ =\ -p(a,b){[a \  _{-\lambda-T} \ b]}
&\ \ \  \text{(skewsymmetry)} \\
\!\!\!\!\!\!\!\!\!
{[a \ _\lambda \  {[b \ _\mu \ c]}]}-
   p(a,b){[b \ _\mu  \ {[a \ _\lambda \  c]}]}\ =\ 
   {[{[a \ _\lambda \  b]} \ _{\lambda+\mu} \ c]}
&\ \ \ \text{(Jacobi identity)}
\end{array}
\end{equation*}
for $a,b,c\in R$.
Here and further  $p(a,b)=(-1)^{p(a)p(b)}$, where $p(a)\in\Z/2\Z$ is the parity of $a$,
and $\otimes$ stands for the tensor product of vector 
spaces over $\C$.
In the skewsymmetry relation, $[a \  _{-\lambda-T} \ b]$ means that we have 
to replace in $[a \  _{\lambda} \ b]$ the indeterminate $\lambda$ with the
operator $(-\lambda-T)$, acting from the left.
% We call \emph{rank} of a Lie conformal algebra its rank as $\C[T]$ module.
For brevity, we will often drop the prefix super in superalgebra or superspace.
\end{definition}

For a Lie conformal algebra we can define a $\C$-bilinear product 
$R\otimes R \rightarrow  R$ for any $n\in \Z_+=\{0,1,2,\cdots\}$, denoted by
$a\otimes b \mapsto a_{(n)} b$ and given by
\begin{equation}\label{nthprod}
[a \ _\lambda \  b]=\sum_{n\in\bZ_+}\lambda^{(n)}a_{(n)}b \ ,
\end{equation}
where we are using the notation: $\lambda^{(n)}:=\frac{\lambda^n}{n!}$. 

\medskip
% Def of vertex algebras

Vertex algebras can be thought of as a special class 
of Lie conformal superalgebras.
\begin{definition}\label{def_va}
A {\itshape vertex algebra} is a pair $(V,\ \vac)$, where
$V$ is a $\Z_2$--graded $\C[T]$--module, called the {\itshape space of states},
and $\vac$ is an element of $V$, called the {\itshape vacuum state},
endowed with two parity preserving operations:
a $\lambda$--bracket $V\otimes V \rightarrow \C[\lambda]\otimes V$
which makes it a Lie conformal superalgebra,
and a {\itshape normally ordered product}
$V\otimes V \rightarrow V$, denoted by $a\otimes b\mapsto :ab:$,
which makes it a unital differential algebra with unity $\vac$ and
derivative $T$.
They satisfy the following axioms
\begin{enumerate}
\item quasi--associativity
$$\begin{array}{c}
\!\!\!\!\!\!\!\!\!\!\!\!\!\!\!\!\!\!\!\!\!\!\!\!\!\!\!\!\!\!
:(:ab:)c:-:a(:bc:): \ 
=\ :\left(\text{{\Large $\int$}}_0^T d\lambda\ a\right)[b \ _\lambda \  c]: \\
\vphantom{\Bigg(}
\!\!\!\!\!\!\!\!\!\!\!\!\!\!\!\!\!\!\!\!\!\!\!\!\!\!\!\!\!\!
+\ p(a,b):\left(\text{{\Large $\int$}}_0^T d\lambda\ b\right)[a \ _\lambda \  c]: \ , 
\end{array}$$
\vspace{-20pt}
\item skewsymmetry of the normally ordered product
$$
:ab:-p(a,b):ba:\ =
\int_{-T}^0 d\lambda [a \ _\lambda \  b] \ ,
$$
\item non commutative Wick formula
$$
[a \ _\lambda \ :bc:]\ =\ 
:[a \ _\lambda \  b]c:+p(a,b):b[a \ _\lambda \  c]:
+\int_0^\lambda d\mu [[a \ _\lambda \  b] \ _\mu \  c] \ .
$$
\end{enumerate}
In quasi--associativity, the first (and similarly the second) integral in the right hand side
should be understood as
$$
\sum_{n\geq0}:(T^{(n+1)}a)b_{(n)}c:\ .
$$
\end{definition}

An equivalent and more familiar definition of vertex algebras
can be given in terms of local fields,  \cite{kac1}.
A proof of the equivalence between the above definition of a vertex algebra
and the one given in \cite{kac1} can be found in \cite{bakalov}.
It also follows from \cite{kac1} that these definitions are equivalent
to the original one of \cite{borcherds}.

\begin{remark}
Since $\vac$ is the unit element of the differential algebra $(V,T)$,
we have $T\vac=0$. Moreover, by sesquilinearity of the $\lambda$--bracket,
it is easy to prove that the torsion of the $\C[T]$--module $V$ is central
with respect to the $\lambda$--bracket;
namely, if $P(T)a=0$ for some $P(T)\in\C[T]\backslash \{0\}$, 
then $[a \ _\lambda \  b]=0$ for all $b\in V$.
In particular the vacuum element is central:
$[a \ _\lambda \  \vac]=[\vac \ _\lambda \  a]=0$, $\forall a\in V$. 
\end{remark}

\begin{remark}
Using skewsymmetry (of both the $\lambda$--bracket and 
the normally ordered product) and non commutative Wick formula, 
one can prove the right Wick formula \cite{bakalov}

$$\begin{array}{c}
\!\!\!\!\!\!\!\!\!\!\!\!\!\!\!\!\!\!\!\!\!\!\!\!
[:ab: \ _\lambda \ c]\ =\ 
:\left(e^{T\partial_\lambda}a\right)[b \ _\lambda \  c]:
+p(a,b):\left(e^{T\partial_\lambda}b\right)[a \ _\lambda \  c]: \\ 
\!\!\!\!\!\!\!\!\!\!\!\!\!\!\!\!\!\!\!\!\!\!\!\!
\vphantom{\Bigg\{}
+\ p(a,b)\text{{\Large $\int$}}_0^\lambda d\mu [b \ _\mu \ [a \ _{\lambda-\mu} \  c]] \ .
\end{array}$$
\end{remark}

For a vertex algebra $V$ we can define $n$--th products for every $n\in\Z$
in the following way.
Since $V$ is a Lie conformal algebra, all $n$-th products with $n\geq0$
are already defined by (\ref{nthprod}).
For $n\leq-1$ we define $n$--th product as
\begin{equation}\label{nthprod2}
a_{(n)}b  = \ :(T^{(-n-1)}a)b: \ .
\end{equation}

% universal enveloping vertex algebra

By definition every vertex algebra is a Lie conformal algebra.
On the other hand, given a Lie conformal algebra $R$ there is a canonical
way to construct a vertex algebra which contains $R$ as a Lie conformal subalgebra.
This result is stated in the following
\begin{theorem}\label{envelopingva}
Let $R$ be a Lie conformal superalgebra with $\lambda$--bracket 
$[a \ _\lambda \  b]$.
Let $R_{\mbox{Lie}}$ be $R$ considered as a Lie superalgebra over $\C$
with respect to the Lie bracket:
$$
[a,b]=
\int_{-T}^0 d\lambda [a \ _\lambda \  b] \ , \qquad a,b\in R \ ,
$$
and let $V=U(R_{\mbox{Lie}})$ be its universal enveloping superalgebra.
Then there exists a unique structure of a vertex superalgebra on $V$
such that the restriction of the $\lambda$--product to 
$R_{\mbox{Lie}}\times R_{\mbox{Lie}}$ coincides 
with the $\lambda$--bracket on $R$
and the restriction of the normally ordered product to 
$R_{\mbox{Lie}}\times V$ coincides with the associative product 
of $U(R_{\mbox{Lie}})$.
\end{theorem}
The vertex algebra thus obtained is denoted by $U(R)$ and is called
the {\itshape universal enveloping vertex algebra} of $R$.
For a proof of this theorem, see \cite{kac1}, \cite{gms},
\cite[Th 7.12]{bakalov}.

% important definitions

In the reminder of this section we introduce some definitions which will be used in the sequel.

\begin{definition}\label{new_free}
Let $R$ be a ($\Z/2\Z$ graded) $\C[T]$--submodule of a vertex algebra $V$.
Choose a basis $\A=\{a_i\ ,\ i\in\I\}$ of $R$ compatible with the $\Z/2\Z$ gradation,
where $\I$ is an ordered set.
\begin{enumerate}
\item One says that $V$ is {\itshape strongly generated} by $R$ if the normally
ordered products of elements of $\A$ span $V$.
\item One says that $V$ is {\itshape freely generated} by $R$ if the set
of elements
$$
\B\ =\
\bigg\{
:a_{i_1} \dots a_{i_n}:
\left|\begin{array}{c}
i_1\leq i_2\leq\dots\leq i_n\ , \\
i_k<i_{k+1} \text{ if } p(a_{i_k})=\bar{1}
\end{array}\right\}
$$
forms a basis of $V$ over $\C$.
In this case $V$ is called a {\itshape freely generated} vertex algebra.
\end{enumerate}
The normally ordered product of more than two elements is defined
by taking products from right to left.
In other words, for elements $a,b,c,\dots\in V$, we will denote
$:abc\dots:\ =\ :a(:b(:c\cdots:):):$.
Also, by convention, the empty normal ordered product is $\vac$, and it is included in $\B$.
\end{definition}

For example, the universal enveloping vertex algebra $U(R)$ of a Lie conformal algebra $R$
is freely generated by $R$, by Theorem \ref{envelopingva}.

From now on, fix an ordered abelian semigroup $\Gamma$ with a zero element $0$,
such that for every $\Delta\in\Gamma$ there are only finitely many elements
$\Delta^\prime\in\Gamma$ such that $\Delta^\prime<\Delta$.
The most important example is a discrete subset of $\R_+$ containing 0
and closed under addition.

\begin{definition}\label{g_gradation}
By a {\itshape $\Gamma$--gradation} of a vector space $U$
we mean a decomposition
$$
U\ =\ \bigoplus_{\Delta\in\Gamma}U[\Delta]\ ,
$$
in a direct sum of subspaces labelled by $\Gamma$.
If $a\in U[\Delta]$, we call $\Delta$ the degree of $a$,
and we will denote it by $\Delta(a)$ or $\Delta_a$.
By convention $0\in U$ is of any degree.
The associated $\Gamma$--filtration of $U$ is defined by letting
$$
U_\Delta\ =\ \bigoplus_{\Delta^\prime\leq\Delta}U[\Delta^\prime]\ .
$$
Let $\T(U)$ be the tensor algebra over $U$, namely
$$
\T(U)\ =\ \C\oplus U\oplus U^{\otimes2}\oplus\cdots\ .
$$
A $\Gamma$--gradation of $U$ extends to a $\Gamma$--gradation of $\T(U)$,
$$
\T(U)\ =\ \bigoplus_{\Delta\in\Gamma}\T(U)[\Delta]\ ,
$$
by letting
$$
\Delta(1)=0\ ,\ \ \Delta(A\otimes B)=\Delta(A)+\Delta(B)\ ,\ \ \forall A,b\in\T(U)\ .
$$
The induced $\Gamma$--filtration is denoted by 
$$
\T_\Delta(U)\ =\ \bigoplus_{\Delta^\prime\leq\Delta}\T(U)[\Delta^\prime]\ .
$$
If moreover $U$ is a $\C[T]$--module, we extend the action of $T$ to $\T(U)$
by Leibniz rule:
$$
T(1)=0\ ,\ \ T(A\otimes B)=T(A)\otimes B+A\otimes T(B)\ ,\ \ \forall A,B\in\T(U)\ .
$$
\end{definition}

%%%%%%%% SECTION 3 %%%%%%%%%%%%%%%%%%%%%%%%%%%%%

\section{Non--linear Lie conformal superalgebras and corresponding 
universal enveloping vertex algebras}
\setcounter{equation}{0}

% introduction

In the previous section we have seen that, to every Lie conformal algebra $R$,
we can associate a universal enveloping vertex algebra $U(R)$, freely generated by $R$.
In general, though, it is not true that a freely generated vertex algebra $V$ is obtained as 
the universal enveloping vertex algebra over a finite Lie conformal algebra.
In fact the generating set $R\subset V$ needs not be closed under the $\lambda$--bracket.

In this section we will introduce the notion of {\itshape non--linear Lie conformal 
superalgebra}, which is  "categorically equivalent" to that 
of freely generated vertex algebra:
to every non--linear Lie conformal algebra $R$ we can associate a universal enveloping
vertex algebra $U(R)$, freely generated by $R$.
Conversely, let $V$ be a vertex algebra freely generated 
by a free $\C[T]$--module $R$ satisfying a "grading condition";
then $R$ is a non--linear Lie conformal algebra and $V$
is isomorphic to the universal enveloping vertex algebra $U(R)$.

% Definition of non-linear conformal algebra

\begin{definition}\label{def1}
A  {\itshape non--linear conformal superalgebra} $R$ is a $\C[T]$--module
which is $\Gamma\backslash\{0\}$--graded by $\C[T]$--submodules:
$R=\oplus_{\Delta\in\Gamma\backslash\{0\}}R[\Delta]$,
and is endowed with a {\itshape $\lambda$--bracket},
namely a parity preserving $\C$--linear map
$$
[\ _\lambda\ ]\ :\ \ R\otimes R\ \longrightarrow\ \C[\lambda]\otimes\T(R)\ ,
$$
such that the following conditions hold
\begin{enumerate}
\item{sesquilinearity}
\begin{eqnarray*}
{[Ta\ _\lambda\ b]} &=& -\lambda{[a\ _\lambda\ b]}\ , \\
{[a\ _\lambda\ Tb]} &=& (\lambda+T){[a\ _\lambda\ b]}\ .
\end{eqnarray*} 
\item{grading condition}
$$
\Delta([a\ _\lambda\ b])\ <\ \Delta(a)+\Delta(b)\ .
$$
\end{enumerate}
\end{definition}
The grading condition can be expressed in terms of the $\Gamma$--filtration of $\T(R)$ 
(see Definition \ref{g_gradation}) in the following way
$$
[R[\Delta_1]\ _\lambda\ R[\Delta_2]]\ \subset\ \C[\lambda]\otimes\T_{\Delta}(R)\ ,
\quad \text{ for some } \Delta<\Delta_1+\Delta_2\ .
$$
Note also that $\T(R)[0]=\C$ and $\T(R)[\delta]=R[\delta]$,
where $\delta$ is the smallest non--zero element of $\Gamma$ such that $R[\delta]\neq0$.
This will be important for induction arguments.

% Claim1:
% (a) maps N and \P_l are well deined on \T_2(R) \X \T(R)
% (b) they respect gradation

The following lemma allows us to introduce the normally ordered product on $\T(R)$
and to extend the $\lambda$--bracket to the whole tensor algebra $\T(R)$, by using
quasi--associativity and the left and right Wick formulas.
\begin{lemma}\label{extension}
Let $R$ be a non--linear conformal algebra. \\
A) There exist unique linear maps
\begin{eqnarray*}
\N\ :\ \  \T(R)\otimes \T(R) &\longrightarrow& \T(R) \ , \\
\P_\lambda\ :\ \ \T(R)\otimes \T(R)
&\longrightarrow& \C[\lambda]\otimes\T(R) \ ,
\end{eqnarray*}
such that, for $a,b,c,\dots\in R,\ A,B,C,\dots\in \T(R)$, the following conditions hold
\begin{eqnarray}
\N(1,A) &=& \N(A,1)\ =\ A\ , \label{1} \\
\N(a,B) &=& a\otimes B\ , \label{2} \\
\N(a\otimes B,C) &=& \N(a,\N(B,C))
 +\N(\Big(\int_0^Td\lambda\ a\Big), \P_\lambda(B,C)) \nonumber\\
&+& p(a,b)\N(\Big(\int_0^Td\lambda\ B\Big)\P_\lambda(a,C))\ , \label{3} \\
\P_\lambda(1,A) &=& \P_\lambda(A,1)\ =\ 0\ ,\label{4} \\
\P_\lambda(a,b) &=& [a\ _\lambda\ b]\ ,\label{5} \\
\P_\lambda(a,b\otimes C) 
&=& \N(\P_\lambda(a,b),C)+p(a,b)\N(b,\P_\lambda(a,C)) \nonumber\\
&+& \int_0^\lambda d\mu\  \P_\mu(\P_{\lambda}(a,b),C) \ , \label{6} \\
\P_\lambda(a\otimes B,C) 
&=& \N(\Big(e^{T\partial_\lambda}a\Big),\P_\lambda(B,C)) 
+ p(a,B)\N(\Big(e^{T\partial_\lambda}B\Big),\P_\lambda(a,C)) \nonumber\\
&+& p(a,B)\int_0^\lambda d\mu\  \P_\mu(B,\P_{\lambda-\mu}(a,C)) \ , \label{7}
\end{eqnarray}
B) These linear maps satisfy the following grading conditions ($A,B\in\T(R)$):
\begin{equation}\label{8}
\Delta(\N(A,B))\leq\Delta(A)+\Delta(B)\ ,\ \ \Delta(\P_\lambda(A,B))<\Delta(A)+\Delta(B)\ .
\end{equation}
\end{lemma}
\begin{proof}
We want to prove, by induction on $\Delta=\Delta(A)+\Delta(B)$, that 
$\N(A,B)$ and $\P_\lambda(A,B)$ exist and are uniquely defined by equations (\ref{1}--\ref{7})
and they satisfy the grading conditions (\ref{8}).
% A
Let us first look at $\N(A,B)$. If $A\in\C$ or $A\in R$, $\N(A,B)$ is defined by conditions 
(\ref{1}) and (\ref{2}).
Suppose then $A=a\otimes A^{\prime}$, where $a\in R$ 
and $A^{\prime}\in R\oplus R^{\otimes2}\oplus\cdots$.
Notice that, in this case, $\Delta(a)<\Delta(A),\ \Delta(A^{\prime})<\Delta(A)$.
By condition (\ref{3}) we have
\begin{eqnarray*}
\N(A,B) &=& a\otimes\N(A^{\prime},B)
 +\N(\Big(\int_0^Td\lambda\ a\Big), \P_\lambda(A^{\prime},B)) \\
&+& p(a,A^{\prime})\N(\Big(\int_0^Td\lambda\ A^{\prime}\Big)\P_\lambda(a,B)\ , 
\end{eqnarray*}
and, by the inductive assumption, each term in the right hand side is uniquely defined
and is in $\T_\Delta(R)$.
% B
Consider then $\P_\lambda(A,B)$. If either $A\in\C$ or $B\in\C$ we have, by (\ref{4}),
$\P_\lambda(A,B)=0$. Moreover, if $A,B\in R$, $\P_\lambda(A,B)$ is defined by (\ref{5}).
Suppose now $A=a\in R,\ B=b\otimes B^{\prime}$, 
with $b\in R$ and $B^{\prime}\in R\oplus R^{\otimes2}\cdots$.
In this case we have, by (\ref{6})
\begin{eqnarray*}
\P_\lambda(A,B) 
&=& \N(\P_\lambda(a,b),B^{\prime})+p(a,b)\N(b,\P_\lambda(a,C)) \\
&+& \int_0^\lambda d\mu\  \P_\mu(\P_{\lambda}(a,b),B^{\prime}) \ ,
\end{eqnarray*}
and each term in the right hand side is well defined by induction, and it lies in
$\T_{\Delta^\prime}(R)$, for some $\Delta^\prime<\Delta$.
Finally, consider the case $A=a\otimes A^{\prime}$, 
with $a\in R$ and $A^{\prime},B\in R\oplus R^{\otimes2}\oplus\cdots$.
In this case we have, by (\ref{7}),
\begin{eqnarray*}
\P_\lambda(A,B) 
&=& \N(\Big(e^{T\partial_\lambda}a\Big),\P_\lambda(A^{\prime},B)) 
+ p(a,A^{\prime})\N(\Big(e^{T\partial_\lambda}A^{\prime}\Big),\P_\lambda(a,B)) \\
&+& p(a,A^{\prime})\int_0^\lambda d\mu\  \P_\mu(A^{\prime},\P_{\lambda-\mu}(a,B)) \ ,
\end{eqnarray*}
and as before each term in the right hand side is defined by induction
and it lies in $\T_{\Delta^\prime}(R)$, for some $\Delta^\prime<\Delta$.
\end{proof}

% categorical stuff

\begin{definition}\label{category}
A {\itshape homomorphism} of non--linear conformal 
algebras $\phi:\ R\rightarrow R^\prime$ is a $\C[T]$--module 
homomorphism preserving the $\Z/2\Z$ and $\Gamma$ gradations
such that the induced map $\T(R)\rightarrow\T(R^\prime)$ is a homomorphism
for the operations $\N$ and $\P_\lambda$.
\end{definition}

% definition od \M(R) and U(R) = T(R) / \M(R)

\begin{definition}
Let $R$ be a non--linear conformal algebra. 
%We define the subspace $\M(R)\subset\T(R)$ by
%$$
%\M(R)=\Span_\C
%\left\{\begin{array}{c}
% A\otimes (b\otimes c-p(b,c)c\otimes b) \otimes D \\
% -A\otimes\N(\Big(\int_{-T}^0 d\lambda\ \P_\lambda(b,c)\Big) ,D) 
% \end{array}\right|
%\left.\begin{array}{c}
% b,c\in R \\
% A,D\in\T(R)
%\end{array}\right\}
%$$
%Furthermore, 
For $\Delta\in\Gamma$, we define the subspace $\M_\Delta(R)\subset\T(R)$ by
$$
\M_\Delta(R)=\Span_\C
\left\{\begin{array}{c}
\!\!\!\!
 A\otimes \Big((b\otimes c-p(b,c)c\otimes b) \otimes D \\
\!\!\!\!
 -\N(\Big(\int_{-T}^0 d\lambda\ \P_\lambda(b,c)\Big) ,D)\Big)
 \end{array}\right|
\left.\begin{array}{c}
\!\!\!\vphantom{\Big(}
 b,c\in R,\ A,D\in\T(R), \\
\!\!\!\vphantom{\Big(}
 A\!\otimes\! b\!\otimes\! c\!\otimes\! D\in\T_\Delta(R)
 \end{array}\right\}
$$
Furthermore, we define
$$
\M(R)\ =\ \bigcup_{\Delta\in\Gamma}\M_\Delta(R)\ .
$$
\end{definition}
\begin{remark}
By definition we clearly have $\M_\Delta(R)\subset\M(R)\cap\T_\Delta(R)$. 
But in general $\M_\Delta(R)$ needs not be equal to $\M(R)\cap\T_\Delta(R)$.
In particular, if $\delta$ is the smallest non--zero element of $\Gamma$ such that
$R[\delta]\neq0$,
then $\M_\delta(R)=0$, since, for $b,c\in R$, $\Delta(b)+\Delta(c)\geq2\delta$,
whereas $\M(R)\cap\T_\delta(R)$ can be a priori non zero.
\end{remark}

% non-linear Lie conformal algebras

\begin{definition}\label{def}
A {\itshape non--linear Lie conformal algebra} is a non--linear conformal algebra $R$
such that the $\lambda$--bracket $[\ _\lambda\ ]$ satisfies the following additional axioms:
\begin{enumerate}
\item skewsymmetry
$$
[a\ _\lambda\ b]\ =\ -p(a,b)[b\ _{-\lambda-T}\ a]\ ,\quad \forall a,b\in R\ ,
$$
\item Jacobi identity
\begin{eqnarray}
\vphantom{\Big(}
&& \P_\lambda(a,\P_\mu(b,c))-p(a,b)\P_\mu(b,\P_\lambda(a,c)) \label{J}\\
\vphantom{\Big(}
&& - \P_{\lambda+\mu}(\P_\lambda(a,b),c) \in \C[\lambda,\mu]\otimes\M_{\Delta^\prime}(R)\ ,\nonumber
\end{eqnarray}
for every $a,b,c\in R$ and for some $\Delta^\prime\in\Gamma$ such that
$\Delta^\prime<\Delta(a)+\Delta(b)+\Delta(c)$.
\end{enumerate}
\end{definition}

% Lie bracket

\begin{remark}\label{liebrack}
Given a $\lambda$--bracket on $R$, consider the map
$\P:\ \T(R)\otimes\T(R)\rightarrow \T(R)$ given by
$$
\P(A,B)=\int_{-T}^0d\lambda\ \P_\lambda(A,B) \ .
$$
It is immediate to show that, if $R$ is a non--linear Lie conformal algebra, 
then $\P$ satisfies skewsymmetry
$$
\P(a,b)=-p(a,b)\P(b,a) \ , 
$$
and Jacobi identity
$$
\P(a,\P(b,c))-p(a,b)\P(b,\P(a,c)) - \P(\P(a,b),c) \in \M_{\Delta^\prime}(R)\ ,
$$
for every $a,b,c\in R$ and for some $\Delta^\prime\in\Gamma$ such that
$\Delta^\prime<\Delta(a)+\Delta(b)+\Delta(c)$.
Thus we obtain a non--linear Lie superalgebra, discussed in the introduction.
\end{remark}

% remark on skew-symmetry axiom

\begin{remark}
In the definition of non--linear Lie conformal algebra we could write the
skew--symmetry axiom in the weaker form:
$$
[a\ _\lambda\ b] + p(a,b)[b\ _{\-\lambda-T}\ a]\ \in\ \M_{\Delta^\prime}(R)\ ,
$$
for some $\Delta^\prime<\Delta(a)+\Delta(b)$.
In this way we would get a similar notion of non--linear Lie conformal algebra,
which, as it will be clear from the results in Section 6,
is not more useful than that given by Definition \ref{def},
from the point of view of classification of vertex algebras.
\end{remark}

% Main Theorem

We are now ready to state the main results of this paper.
\begin{theorem}\label{main}
Let $R$ be a non--linear Lie conformal algebra.
Consider the vector space $U(R)=\T(R)/\M(R)$, 
and denote by $\pi:\ \T(R)\rightarrow U(R)$ the
quotient map, and by $:\ \ :$ the image in $U(R)$ of the tensor product
of elements of $R$:
$$
:ab\cdots c:\ =\ \pi(a\otimes b\otimes \cdots\otimes c)\ ,\quad \forall a,b,\cdots,c\in R\ .
$$
Let $\A=\{a_i,\ i\in\II\}$ be an ordered basis of $R$
compatible with the $\Z/2\Z$--gradation and the $\Gamma$--gradation.
We denote by $\B$ the image in $U(R)$ of the collection of all ordered monomials,
namely
$$
\B=\left\{\begin{array}{c}
:a_{i_1}\dots a_{i_n}:\ \Big|\ i_1\leq i_2\leq\dots\leq i_n\ , \\
\text{ and } i_k<i_{k+1} \text{ if } p(a_{i_k})=\bar{1}
\end{array}\right\}\subset U(R)\ .
$$
Then
\begin{enumerate}
\item $\B$ is a basis of the vector space $U(R)$. In particular we have a natural embedding
$$
R\ \stackrel{\sim}{\rightarrow}\ \pi(R)\ \subset\ U(R)\ .
$$
\item There is a canonical structure of a vertex algebra on $U(R)$, called
the {\itshape universal enveloping vertex algebra of $R$}, such that the vacuum vector $\vac$ is $\pi(1)$,
the infinitesimal translation operator $T:\ U(R)\rightarrow U(R)$ is induced by the action
of $T$ on $\T(R)$:
$$
T(\pi(A))\ =\ \pi(TA)\ ,\quad \forall A\in\T(R)\ ,
$$
the normally ordered product on $U(R)$ is induced by $\N$:
$$
:\pi(A)\pi(B):\ =\ \pi(\N(A,B))\ ,\quad \forall A,B\in\T(R) \ ,
$$
and the $\lambda$--bracket $[\ _\lambda\ ]:\ U(R)\otimes U(R)\rightarrow \C[\lambda]U(R)$
is induced by $\P_\lambda$:
$$
[\pi(A)\ _\lambda\ \pi(B)]\ =\ \pi(\P_\lambda(A,B))\ ,
\quad \forall A,B\in\T(R) \ .
$$
\end{enumerate}
In other words, $U(R)$ is a vertex algebra freely generated by $R$.
\end{theorem}

% remark: universal property

\begin{remark}\label{univ}
The vertex algebra $U(R)$ has the following universal property:
let $\phi$ be a homomorphism of $R$ to a pre--graded vertex algebra $V$ (see Definition \ref{qgrad}),
i.e. a $\C[T]$--module homomorphism preserving $\Z/2\Z$ and $\Gamma$--gradations
such that the induced map $\T(R)\rightarrow V$ is a homomorphism for 
the operations $\N$ and $\P_\lambda$.
Then there exists a unique vertex algebra homomorphism 
$\phi^\prime:\ U(R)\rightarrow V$ such that $\phi=\phi^\prime\circ\pi$.
\end{remark}

% remark: independence of the choice of grading

\begin{remark}
Notice that the universal enveloping vertex algebra $U(R)$ is independent on the choice 
of the grading of $R$.
\end{remark}

%%% remark: case R is a conformal algebra

\begin{example}
% Lie conf alg ==> non-linear Lie conf alg
Suppose $\hat{R}=R\oplus\C\vac$, where $\C\vac$ is a torsion submodule,
is a Lie conformal algebra.
% 1
In this case, if we let $\Gamma=\Z_+$ and
assign degree $\Delta=1$ to every element of $R$, we make $R$ into
a non--linear Lie conformal algebra.
% 2
In this case we have a Lie algebra structure on $R$ defined by the Lie bracket
(recall that $\C\vac$ is in the center of the $\lambda$--bracket)
$$
[a,b]\ =\ \int_{-T}^0 d\lambda[a\ _\lambda\ b]\ ,\quad \forall a,b\in R\ .
$$
Therefore the space $U(R)$ coincides with the universal enveloping algebra of $R$ 
(viewed as a Lie algebra with respect to $[\ , \ ]$), and the first part of Theorem \ref{main}
is the PBW theorem for ordinary Lie superalgebras.
% 3
Moreover in this case $U(R)$ simply coincides with the quotient of 
the universal enveloping vertex algebra of $\hat{R}$, 
defined by Theorem \ref{envelopingva}, by the ideal generated by $\vac-1$.
In particular, if $\hat{R}$ is the Virasoro Lie conformal algebra
$[L\ _\lambda\ L]\ =\ (T+2\lambda)L+\frac{c}{12}\lambda^3\vac$,
then $U(R)$ is the universal Virasoro vertex algebra
with central charge $c$.

% non-linear Lie conf alg with degree 1 ==> Lie conf alg
Conversely, suppose $R$ is a non--linear Lie onformal algebra 
graded by $\Gamma=\Z_+$ and such that every element of $R$ has degree $\Delta=1$. 
It follows by the grading condition on $[\ _\lambda\ ]$ that 
$\hat{R}=R\oplus\C$ is a Lie conformal algebra, and therefore Theorem \ref{main}
holds by the above considerations.
\end{example}

% Example: Zamolodchikov

\begin{example}
The first example of a "genuinely" non--linear Lie conformal algebra $R$ is
Zamolodchikov's $W_3$--algebra \cite{zam}, which, in our language, is
$$
W_3\ =\ \C[T]L+\C[T]W\ ,
$$
where $\Delta(L)=2,\ \Delta(W)=3$, and the $\lambda$--brackets are as follows:
\begin{eqnarray}
\vphantom{\Big(}
[L\ _\lambda\ L] &=& (T+2\lambda)L+\frac{c}{12}\lambda^3\ ,
\qquad [L\ _\lambda\ W]\ =\ (T+3\lambda)W\ ,\nonumber\\
\vphantom{\Big(}
[W\ _\lambda\ W] &=& (T+2\lambda)\bigg(
\frac{16}{22+5c}(L\otimes L)
+\frac{c-10}{3(22+5c)}T^2L \label{zeq}\\
&+&\frac{1}{6}\lambda(T+\lambda)L\bigg)
+\frac{c}{360}\lambda^5\ .\nonumber
\end{eqnarray}
We want to prove that this is the unique, up to isomorphisms, non--linear Lie conformal algebra
generated by two elements $L$ and $W$, such that $L$ is a Virasoro element of central charge $c$
and $W$ is an even primary element of conformal weight 3.
We assume that $[W\ _\lambda\ W]\not\in\C[\lambda]1$.

%The above assumptions fix uniquely the $\lambda$--brackets
%$[L\ _\lambda\ L]$ and $[L\ _\lambda\ W]$ as in (\ref{zeq}),
%and by skewsymmetry we also get $[W\ _\lambda\ L]$.
We only need to show that $[W\ _\lambda\ W]$ has to be as in (\ref{zeq}).
By simple conformal weight considerations (see Example \ref{cw})
and by skewsymmetry of the $\lambda$--bracket,
the most general form of $[W\ _\lambda\ W]$ is
$$
[W\ _\lambda\ W] = (T+2\lambda)\Big(
\alpha(L\otimes L)+\beta TW + \gamma T^2L
+ \delta\lambda(T+\lambda)L\Big)
+\epsilon\lambda^5\ ,
$$
for $\alpha,\beta,\gamma,\delta,\epsilon\in\C$.
We need to find all possible values of these parameters such that the
Jacobi identity (\ref{J}) holds.
We may assume $\alpha\neq0$, since otherwise $W_3\oplus\C\vac$
is a "linear" Lie conformal algebra, and from their classification
\cite{dandrea1} it follows that $[W _\lambda\ W]\in\C[\lambda]1$, 
which is not allowed.
There are only two non trivial Jacobi identities (\ref{J}), namely the ones involving
two or three $W$ elements.
We will show how to deal with the first, and we will leave the second 
to the reader.
We need to impose
\begin{equation}\label{jac_w3}
[W\ _\lambda\ [W\ _\mu\ L]]-[W\ _\mu\ [W\ _\lambda\ L]]\ 
\equiv\ [[W\ _\lambda\ W]\ _{\lambda+\mu}\ L]\quad \mod\C[\lambda,\mu]\otimes\M_7(R)\ .
\end{equation}
Using the axioms of non--linear conformal algebra,
we can compute every term of equation (\ref{jac_w3}).
The first term of the left hand side is
\begin{eqnarray}\label{w1}
\vphantom{\Big(}&
(2T+2\lambda+3\mu)(T+2\lambda)\Big\{\alpha L\otimes L+\beta TW
+\gamma T^2\lambda 
&\nonumber\\&\vphantom{\Big(}
+\delta\lambda(T+\lambda)L\Big\}
+(2T+2\lambda+3\mu)\epsilon\lambda^5\ ,
\end{eqnarray}
and the second term is obtained by replacing $\lambda$ by $\mu$.
The right hand side of (\ref{jac_w3}) is
\begin{eqnarray}\label{w2}
&(\lambda-\mu)\bigg\{
2\alpha L\otimes TL
+4\alpha TL\otimes L
+4\alpha (\lambda+\mu)L\otimes L 
+\frac{1}{6}\alpha c(T+\lambda+\mu)^3L \nonumber\\
&+\alpha\Big(
	\frac{4}{3}(\lambda+\mu)^3
	+\frac{5}{2}(\lambda+\mu)^2T
	+(\lambda+\mu)T^2\Big)L
+\frac{1}{40}\alpha c(\lambda+\mu)^5 \nonumber\\
&-\beta(\lambda+\mu)(2T+3\lambda+3\mu)W \\
&+(\gamma(\lambda+\mu)^2-\delta\lambda\mu)
	((T+2\lambda+2\mu)L+\frac{c}{12}(\lambda+\mu)^3)
\bigg\}\ . \nonumber
\end{eqnarray}
If $d\neq0$, (\ref{jac_w3}) can't be an exact identity, 
but only an identity modulo $\M_\Delta(R)$.
Indeed in (\ref{w2}) we have a term $2\alpha L\otimes TL$,
which does not appear in (\ref{w1}).
But we can replace in (\ref{w2})
$$
TL\otimes L-L\otimes TL\quad \text{ by }\quad \frac{1}{6} T^3L\quad \mod\M_4(R)\ .
$$
After this substitution,
both sides of (\ref{jac_w3}) are linear combinations 
of $L\otimes L,\ L,\ W,\ 1$, with coefficients in $\C[\lambda,\mu,T]$.
Hence by Theorem \ref{main} all these coefficients are zero.
It is then a simple exercise
to show that the only solution of (\ref{jac_w3})
is given, up to rescaling of $W$, by choosing the parameters
$\alpha,\beta,\gamma,\delta,\epsilon$ as in (\ref{zeq}).
After this, one checks that the Jacobi identity involving 
three $W$ elements is automatic.
\end{example}

% quantum reduction

A large class of non--linear Lie conformal (super)algebras is obtained by quantum
Hamiltonian reduction attached to a simple Lie (super)algebra $\g$ and a nilpotent orbit
in $\g$ (see e.g. \cite{frenkel2},\cite{kac_wak_2},\cite{bt}).
In particular, the Virasoro algebra and the $W_3$--algebra 
are obtained by making use of the principal
nilpotent orbit of $\g=\ssl_2$ and $\ssl_3$ respectively.

% proofs in the following.

The next two sections will be devoted to the proof Theorem \ref{main}.
In the next section we will prove some technical results,
and in the following section we will complete the proof of Theorem \ref{main}.

%%%%%%%% SECTION 4 %%%%%%%%%%%%%%%%%%%%%%%%%%%%%%%%

\section{Technical results}
\setcounter{equation}{0}

% - notation

Let $R$ be a non--linear Lie conformal algebra. 
For convenience, we introduce the following notations which will be used
throughout the following sections
($A,B,C\dots\in\T(R)$):
$$\begin{array}{rcl}
\skewl(A,B;\lambda) & = & 
 \P_\lambda(A,B)\ +\ p(A,B)\P_{-\lambda-T}(B,A) \ , \\
\vphantom{\Bigg(}
\skewn(A,B,C) & = &
 \N(A,\N(B,C))\ -\ p(A,B)\N(B,\N(A,C)) \\
 & - & \N(\Big(\int_{-T}^0d\lambda\ \P_\lambda(A,B)\Big),C) \ , \\
\vphantom{\Bigg(}
\lw(A,B,C;\lambda) & = &
 \P_\lambda(A,\N(B,C))\ -\ \int_0^\lambda d\mu\ \ \P_\mu(\P_\lambda(A,B),C) \\
 & - & \N(\P_\lambda(A,B),C)\ -\ p(A,B)\N(B,\P_\lambda(A,C)) \ , \\
\vphantom{\Bigg(}
\rw(A,B,C;\lambda) & = &
 \P_\lambda(\N(A,B),C)\ 
 -\ p(A,B)\int_0^\lambda d\mu\ \P_\mu(B,\P_{\lambda-\mu}(A,C)) \\
 & - & \N(\Big(e^{T\partial_\lambda}A\Big),\P_\lambda(B,C))\ 
 -\ p(A,B)\N(\Big(e^{T\partial_\lambda}B\Big),\P_\lambda(A,C)) \ , \\
\vphantom{\Bigg(}
\qa(A,B,C) & = &
 \N(\N(A,B),C)\ -\ \N(A,\N(B,C))\ \\
 & - &  \N(\Big(\int_0^T d\lambda\ A\Big),\P_\lambda(B,C)) \\
\vphantom{\Bigg(}
 & - & p(A,B)\N(\Big(\int_0^T d\lambda\ B\Big),\P_\lambda(A,C)) \ , \\
\j(A,B,C;\lambda,\mu) & = &
 \P_\lambda(A,\P_\mu(B,C))\ -\ p(A,B)\P_\mu(B,\P_\lambda(A,C)) \\ 
\vphantom{\Bigg(}
 & - & \P_{\lambda+\mu}(\P_\lambda(A,B),C) \ .
\end{array}$$

% - remark: rewrite the definitions

\begin{remark}
Using the above notation, we can write in a more concise form all the
definitions introduced in the previous section.
The maps $\N:\ \T(R)\otimes \T(R)\rightarrow \T(R)$ and
$\P_\lambda:\ \T(R)\otimes \T(R)\rightarrow \C[\lambda]\otimes\T(R)$
are defined respectively by the equations
$$
\qa(a,B,C)\ =\ 0\ ,\quad \lw(a,b,C;\lambda)\ =\ 0\ ,\quad \rw(a,B,C;\lambda)\ =\ 0\ , 
$$
for all $a,b\in R$ and $B,C\in\T(R)$ .
The axioms of non--linear Lie conformal algebras can be written as 
$$
\skewl(a,b\ ;\lambda) = 0\ ,
\quad
\j(a,b,c\ ;\lambda,\mu) \in \C[\lambda,\mu]\otimes\M_{\Delta^\prime}(R)\ ,
$$
for all $a,b,c\in R$ and for some $\Delta^\prime<\Delta(a)+\Delta(b)+\Delta(c)$.
Finally, the subspaces $\M_\Delta(R)\subset\T(R),\ \Delta\in\Gamma$, are defined as
$$
\M_\Delta(R)=\Span_\C
\left\{
\vphantom{\bigg(}
A\otimes\skewn(b,c,D) 
\right|
\left.\begin{array}{c}
A,D\in\T(R),\ b,c\in R\ , \\
\Delta(A)+\Delta(b)+\Delta(c)+\Delta(D)\leq\Delta
\end{array}\right\}\ ,
$$
and, by definition, $\M(R)=\cup_{\Delta\in\Gamma}\M_\Delta(R)$.
\end{remark}

% - lemma 0: sesquilinearity and leibniz rule

\begin{lemma}
The endomorphism $T:\ \T(R)\rightarrow\T(R)$ is a derivation of the product 
$\N:\ \T(R)\otimes\T(R)\rightarrow\T(R)$, namely, for $A,B\in\T(R)$,
\begin{equation}\label{leib}
T\N(A,B)=\N(T A,B)+\N(A,TB)\ .
\end{equation}
Furthermore, the $\lambda$--bracket 
$\P_\lambda:\ \T(R)\otimes \T(R)\rightarrow \C[\lambda]\otimes\T(R)$
satisfies sesquilinearity, namely, for $A,B\in\T(R)$,
\begin{eqnarray}
\P_\lambda(TA,B) &=& -\lambda\P_\lambda(A,B)\ , \label{sesq1}\\
 \P_\lambda(A,TB) &=& (\lambda+T)\P_\lambda(A,B)\ . \label{sesq2}
\end{eqnarray}
In particular $T$ is a derivation of $\P_\lambda$.
\end{lemma}
\begin{proof}
We will prove both statements of the lemma by induction on $\Delta=\Delta(A)+\Delta(B)$.
If $\Delta=0$, we have $A,B\in\C$, and there is nothing to prove.
Suppose then $\Delta>0$. 
For $A\in\C$, all three equations (\ref{leib}), (\ref{sesq1}) and (\ref{sesq2}) are obviously satisfied.
We will consider separately two cases: 
\begin{enumerate}
\item $A=a\in R$,
\item $A=a\otimes A^{\prime}$, 
with $a\in R,\ A^{\prime}\in R\oplus R^{\otimes2}\oplus\cdots\subset\T(R)$.
\end{enumerate}

\noindent {\itshape Case 1.}
For $A\in R$, equation (\ref{leib}) follows by the fact that $T$ is a derivation of the tensor product
and that $N(a,B)=a\otimes B$.
Consider then equations (\ref{sesq1}) and (\ref{sesq2}).
For $B\in\C$ they are trivial, and for $B\in R$ they hold by (\ref{5}).
Suppose then $B=b\otimes B^{\prime}$, with $b\in R$ 
and $B^{\prime}\in R\oplus R^{\otimes2}\oplus\cdots$.
By (\ref{6}) we have
\begin{eqnarray*}
\P_\lambda(Ta,B) 
&=& \N(\P_\lambda(Ta,b),B^{\prime})+p(a,b)\N(b,\P_\lambda(Ta,B^{\prime})) \\
&+& \int_0^\lambda d\mu\  \P_\mu(\P_{\lambda}(Ta,b),B^{\prime}) 
\ =\ -\lambda\P_\lambda(a,B)\ .
\end{eqnarray*}
In the last identity we used inductive assumption and the grading conditions (\ref{8}). 
Similarly we have
\begin{eqnarray*}
\P_\lambda(a,TB) &=& \P_\lambda(a,Tb\otimes B^{\prime}) + \P_\lambda(a,b\otimes TB^{\prime}) \\
&=& \N(\P_\lambda(a,Tb),B^{\prime})+p(a,b)\N(Tb,\P_\lambda(a,B^{\prime})) \\
&+& \int_0^\lambda d\mu\  \P_\mu(\P_{\lambda}(a,Tb),B^{\prime}) \\
&+& \N(\P_\lambda(a,b),TB^{\prime})+p(a,b)\N(b,\P_\lambda(a,TB^{\prime})) \\
&+& \int_0^\lambda d\mu\  \P_\mu(\P_{\lambda}(a,b),TB^{\prime}) 
\ =\ (\lambda+T)\P_\lambda(a,B)\ .
\end{eqnarray*}

\noindent {\itshape Case 2.}
By equation (\ref{3}), inductive assumption and the grading conditions (\ref{8}) we have
\begin{eqnarray*}
T\N(A,B) &=& T\Big\{
\N(a,\N(A^{\prime},B))
+\N(\Big(\int_0^Td\lambda\ a\Big), \P_\lambda(A^{\prime},B)) \\
&+& p(a,A^{\prime})\N(\Big(\int_0^Td\lambda\ A^{\prime}\Big),\P_\lambda(a,B))
\Big\}
\ =\ \N(TA,B)+\N(A,TB)\ ,
\end{eqnarray*}
which proves (\ref{leib}). Similarly by equation (\ref{7}) we prove (\ref{sesq1}) and (\ref{sesq2}):
\begin{eqnarray*}
\P_\lambda(TA,B) &=&  \P_\lambda(Ta\otimes A^{\prime},B)+\P_\lambda(a\otimes TA^{\prime},B) \\
&=& \N(\Big(e^{T\partial_\lambda}Ta\Big),\P_\lambda(A^{\prime},B)) 
+ p(a,A^{\prime})\N(\Big(e^{T\partial_\lambda}A^{\prime}\Big),\P_\lambda(Ta,B)) \\
&+& p(a,A^{\prime})\int_0^\lambda d\mu\  \P_\mu(A^{\prime},\P_{\lambda-\mu}(Ta,B)) \\
&+& \N(\Big(e^{T\partial_\lambda}a\Big),\P_\lambda(TA^{\prime},B)) 
+ p(a,A^{\prime})\N(\Big(e^{T\partial_\lambda}TA^{\prime}\Big),\P_\lambda(a,B)) \\
&+& p(a,A^{\prime})\int_0^\lambda d\mu\  \P_\mu(TA^{\prime},\P_{\lambda-\mu}(a,B))
\ =\ -\lambda\P_\lambda(A,B)\ .
\end{eqnarray*}
and
\begin{eqnarray*}
\P_\lambda(A,TB) 
&=& \N(\Big(e^{T\partial_\lambda}a\Big),\P_\lambda(A^{\prime},TB)) 
+ p(a,A^{\prime})\N(\Big(e^{T\partial_\lambda}A^{\prime}\Big),\P_\lambda(a,TB)) \\
&+& p(a,A^{\prime})\int_0^\lambda d\mu\  \P_\mu(A^{\prime},\P_{\lambda-\mu}(a,TB))
\ =\ (\lambda+T)\P_\lambda(A,B)\ .
\end{eqnarray*}
In the above equations we have used the obvious commutation relation
$e^{T\partial_\lambda}\lambda=(\lambda+T)e^{T\partial_\lambda}$.
This completes the proof of the lemma.
\end{proof}

% - corollary 0: M(R) is T-invariant

As an immediate consequence of the above lemma, we get the following
\begin{corollary}
The subspaces $\M_\Delta(R)\subset\T(R)$ are invariant under the action of $T$.
In particular $T\M(R)\subset\M(R)$.
\end{corollary}

% - lemma 1: all relations

The following lemma follows by a straightforward but quite lengthy computation, which we omit.
\begin{lemma}
The following equations hold for all $a,b,\dots\in R$ and $A,B,\dots\in\T(R)$:
\end{lemma}
% a1
\begin{eqnarray}
\N(\skewn(a,b,C),D) &=& \skewn(a,b,\N(C,D)) 
 + \int_0^{T_a+T_b}d\lambda\ \skewn(a,b,\P_\lambda(C,D)) \nonumber\\
&-& \qa(\Big(\int_{-T}^0d\lambda\ \P_\lambda(a,b)\Big),C,D) \label{a1}\\
&-& p(a,C)p(b,C)\N(\Big(\int_0^Td\lambda\ \int_0^{T-\lambda}d\mu\ C\Big),\j(a,b,D;\lambda,\mu))
\nonumber
\end{eqnarray}
% a2
\begin{eqnarray}
\vphantom{\bigg(}
\P_\lambda(a,\skewn(b,c,D))
&=& \skewn(\P_\lambda(a,b),c,D)-p(b,c)\skewn(\P_\lambda(a,c),b,D) \label{a2}\\
\vphantom{\Big(}
&&\ \ +\ p(a,b)p(a,c)\skewn(b,c,\P_\lambda(a,D)) \nonumber\\
&+& \int_0^\lambda d\mu\Big\{\lw(\P_\lambda(a,b),c,D;\mu)-p(b,c)\lw(\P_\lambda(a,c),b,D;\mu)\Big\} \nonumber\\
&-& \lw(a,\Big(\int_{-T}^0d\mu\ \P_\mu(b,c)\Big),D;\lambda) \nonumber\\
&+& \N(\Big(\int_{-T}^\lambda d\mu 
\Big\{\skewl(\P_\lambda(a,b),c;\mu)- p(b,c)\skewl(\P_\lambda(a,c),b;\mu)\Big\}\Big),D) \nonumber\\
&-& \N(\Big(\int_{-\lambda-T}^0d\mu\  \skewl(a,\P_\mu(b,c);\lambda)\Big),D) \nonumber\\
&-& p(a,b)p(a,c)\N(\Big(\int_{-T}^\lambda d\mu\ \j(b,c,a;-\mu-T,\mu-\lambda)\Big),D) \nonumber\\
&+& \int_0^\lambda d\nu \int_\nu^\lambda d\mu\  \P_\nu(
\skewl(\P_\lambda(a,b),c;\mu) - p(b,c)\skewl(\P_\lambda(a,c),b;\mu),D) \nonumber\\
&-& \int_0^\lambda d\nu \int_\nu^\lambda d\mu\  \P_\nu(\skewl(a,\P_{\mu-\lambda}(b,c);\lambda),D) \nonumber\\
&-& p(a,b)p(a,c)\int_0^\lambda d\nu \int_\nu^\lambda d\mu\  \P_\nu(\j(b,c,a;\mu-\lambda,\nu-\mu),D)
\nonumber
\end{eqnarray}
% a3
\begin{eqnarray}
\P_\lambda(\skewn(a,b,C),D) 
&=& \skewn(\Big(e^{T\partial_\lambda}a\Big),\Big(e^{T\partial_\lambda}b\Big),\P_\lambda(C,D)) 
\label{a3}\\
&-& \rw(\Big(\int_{-T}^0 d\mu\ \P_\mu(a,b)\Big),C,D;\lambda) \nonumber\\
&-& p(a,C)p(b,C)\N(\Big(\int_0^{\lambda+T}d\mu\  e^{T\partial_\lambda} C\Big),\j(a,b,D;\mu,\lambda-\mu)) \nonumber\\
&-& p(a,C)p(b,C)\int_0^\lambda d\mu\int_0^\mu d\nu\  \P_\nu(C,\j(a,b,D;\mu-\nu,\lambda-\mu))
\nonumber
\end{eqnarray}
% a4
\begin{eqnarray}
\qa(\N(a,A^{\prime}),B,C)
&=& \N(a,\qa(A^{\prime},B,C))
+ p(a,A^{\prime})\int_0^{T_{A^{\prime}}}d\lambda\  \qa(A^{\prime},\P_\lambda(a,B),C) \nonumber\\
&-& \int_0^{T_a}d\lambda\  \N(a,\lw(A^{\prime},B,C;\lambda)) \label{a4}\\
&-& p(a,A^{\prime})\int_0^{T_{A^{\prime}}}d\lambda\  \N(A^{\prime},\lw(a,B,C;\lambda)) \nonumber\\
&+& \int_0^{T_a}d\lambda\  \N(a,\rw(A^{\prime},B,C;\lambda)) \nonumber\\
&+& p(a,A^{\prime})p(a,B)\int_0^{T_{A^{\prime}}+T_B}d\lambda\  \qa(A^{\prime},B,\P_\lambda(a,C)) 
\nonumber\\
&+& p(A^{\prime},B)\int_{T_a}^{T_a+T_B}d\lambda\ \skewn(a,B,\P_\lambda(A^{\prime},C)) 
\nonumber\\
&+& p(a,A^{\prime})p(a,B) \int_{T_{A^{\prime}}}^{T_{A^{\prime}}+T_B}d\lambda\ \skewn(A^{\prime},B,\P_\lambda(a,C)) 
\nonumber\\
&-& \int_0^{T_a}d\lambda\ \int_0^{T_a-\lambda}d\mu\ \N(a,\j(A^{\prime},B,C;\lambda,\mu)) 
\nonumber\\
&-& p(a,A^{\prime})\int_0^{T_{A^{\prime}}}d\lambda\ \int_0^{T_{A^{\prime}}-\lambda}d\mu\ \N(A^{\prime},\j(a,B,C;\lambda,\mu)) \nonumber
\end{eqnarray}
% a5
\begin{eqnarray}
\lw(a,\N(b,B^{\prime}),C;\lambda)
&=& p(a,b) \N(b,\lw(a,B^{\prime},C;\lambda)) - \qa(\P_\lambda(a,b),B^{\prime},C) \nonumber\\
&+& \int_0^\lambda d\mu\  \lw(\P_\lambda(a,b),B^{\prime},C;\mu) \label{a5}\\
&+& p(b,B^{\prime})\lw(a,\Big(\int_0^Td\mu\ B^{\prime}\Big),\P_\mu(b,C);\lambda) \nonumber\\
&-& \int_0^\lambda d\mu\  \rw(\P_\lambda(a,b),B^{\prime},C;\mu) \nonumber\\
&+&p(a,b)\N(\Big(\int_0^T d\mu\ b\Big),\j(a,B^{\prime},C;\lambda,\mu)) \nonumber\\
&+& p(a,B^{\prime})p(b,B^{\prime})\N(\Big(\int_0^T d\mu\ B^{\prime}\Big),\j(a,b,C;\lambda,\mu)) 
\nonumber\\
&+& \int_0^\lambda d\mu\int_0^{\lambda-\mu}d\nu\ \j(\P_\lambda(a,b),B^{\prime},C;\mu,\nu)
\nonumber
\end{eqnarray}
% a6
\begin{eqnarray}
\lw(\N(a,A^{\prime}),B,C;\lambda)
&=&  \N(\Big(e^{T\partial_\lambda}a\Big),\lw(A^{\prime},B,C;\lambda)) %\\
+ p(a,A^{\prime})\N(\Big(e^{T\partial_\lambda}A^{\prime}\Big),\lw(a,B,C;\lambda)) \nonumber\\
&-& p(a,A^{\prime}) \qa(\Big(e^{T\partial_\lambda}A^{\prime}\Big),\P_\lambda(a,B),C) %\nonumber\\
+ p(A^{\prime},B)\skewn(\Big(e^{T\partial_\lambda}a\Big),B,\P_\lambda(A^{\prime},C)) \nonumber\\
&+& p(a,A^{\prime})p(a,B)\skewn(\Big(e^{T\partial_\lambda}A^{\prime}\Big),B,\P_\lambda(a,C)) 
\nonumber\\
&+& p(a,A^{\prime})\int_0^\lambda d\mu\  \P_\mu(A^{\prime},\lw(a,B,C;\lambda-\mu)) 
\label{a6}\\
&-& p(a,A^{\prime})\int_0^\lambda d\mu\  \rw(\Big(e^{T\partial_\lambda}A^{\prime}\Big),\P_\lambda(a,B),C;\mu) \nonumber\\
&+& p(a,A^{\prime}) \int_0^\lambda d\mu\  \lw(A^{\prime},\P_{\lambda-\mu}(a,B),C;\mu) \nonumber\\
&+& p(a,A^{\prime})p(a,B) \int_0^\lambda d\mu\  \lw(A^{\prime},B,\P_{\lambda-\mu}(a,C);\mu) 
\nonumber\\
&+& p(a,A^{\prime})\int_0^\lambda d\mu\int_0^{\lambda-\mu}d\nu\  \j(A^{\prime},\P_{\lambda-\mu}(a,B),C;\mu,\nu) \nonumber
\end{eqnarray}
% a7
\begin{eqnarray}
\rw(\N(a,A^{\prime}),B,C;\lambda)
&=& \N(\Big(e^{T\partial_\lambda}a\Big),\rw(A^{\prime},B,C;\lambda)) \label{a7}\\
&+& p(a,A^{\prime})p(a,B)\qa(\Big(e^{T\partial_\lambda}A^{\prime}\Big),
  \Big(e^{T\partial_\lambda}B\Big),\P_\lambda(a,C)) \nonumber\\
&+& p(A^{\prime},B)\skewn(\Big(e^{T\partial_\lambda}a\Big),
  \Big(e^{T\partial_\lambda}B\Big),\P_\lambda(A^{\prime},C)) \nonumber\\
&+& p(a,A^{\prime})p(a,B)\skewn(\Big(e^{T\partial_\lambda}A^{\prime}\Big),
  \Big(e^{T\partial_\lambda}B\Big),\P_\lambda(a,C)) \nonumber\\
&+& p(a,A^{\prime}) \rw(\Big(\int_0^Td\mu\ A^{\prime}\Big),\P_\mu(a,B),C;\lambda) \nonumber\\
&-& p(a,B)p(A^{\prime},B)\int_0^\lambda d\mu\  \lw(B,\Big(e^{T\partial_\lambda}a\Big),
 \P_{\lambda-\mu}(A^{\prime},C);\mu) \nonumber\\
&-& p(a,A^{\prime})p(a,B)p(A^{\prime},B)\int_0^\lambda d\mu\  
  \lw(B,\Big(e^{T\partial_\lambda}A^{\prime}\Big),\P_{\lambda- \mu}(a,C);\mu) \nonumber\\
&+& p(a,A^{\prime})p(a,B)\int_0^\lambda d\mu\  \rw(A^{\prime},B,\P_{\lambda-\mu}(a,C);\mu) \nonumber\\
&-& \N(\Big(\int_0^T d\mu\  e^{T\partial_\lambda}a\Big),\j(A^{\prime},B,C;\mu,\lambda-\mu)) \nonumber\\
&-& p(a,A^{\prime})\N(\Big(\int_0^T d\mu\  e^{T\partial_\lambda}A^{\prime}\Big),
  \j(a,B,C;\mu,\lambda-\mu)) \nonumber\\
&-& p(a,A^{\prime})p(a,B)p(A^{\prime},B)\int_0^\lambda d\mu\ 
 \N(\Big(e^{T\partial_\lambda}\skewl(B,A^{\prime};\mu)\Big),\P_{\lambda}(a,C)) \nonumber\\
&-& p(a,B)p(A^{\prime},B)\int_0^\lambda d\mu\  \N(\Big(e^{T\partial_\lambda}\skewl(B,a;\mu)\Big),
 \P_{\lambda}(A^{\prime},C)) \nonumber\\
&+& p(A^{\prime},B)\int_0^{T_{A^{\prime}}}d\mu\int_0^\lambda d\nu\  \P_\nu(\skewl(a,B;\mu),\P_{\lambda-\nu}(A^{\prime},C)) \nonumber\\
&+& p(a,A^{\prime})p(a,B)\int_0^{T_a}d\mu\int_0^\lambda d\nu\  \P_\nu(\skewl(A^{\prime},B;\mu),\P_{\lambda-\nu}(a,C)) \nonumber
\end{eqnarray}
% a8
\begin{eqnarray}
\j(a,b,\N(c,D);\lambda,\mu)
&=& \lw(a,\P_\mu(b,c),D;\lambda) - p(a,b)\lw(b,\P_\lambda(a,c),D;\mu) \nonumber\\
\vphantom{\Bigg(}
&-& \lw(\P_\lambda(a,b),c,D;\lambda+\mu) \label{a8}\\
&+& \N(\j(a,b,c;\lambda,\mu),D) + p(a,c)p(b,c)\N(c,\j(a,b,D;\lambda,\mu)) \nonumber\\
&+& \int_0^\mu d\nu\  \j(a,\P_\mu(b,c),D;\lambda,\nu)
 - p(a,b)\int_0^\lambda d\nu\  \j(b,\P_\lambda(a,c),D;\mu,\nu) \nonumber\\
 &+& \int_0^{\lambda+\mu}d\nu\  \P_\nu(\j(a,b,c;\lambda,\mu),D) \nonumber
\end{eqnarray}
% a9
\begin{eqnarray}
\vphantom{\Big(}
\j(A,\N(b,B^{\prime}),D;\lambda,\mu)
&=&  \lw(A,b,\P_{\mu+T_b}(B^{\prime},D);\lambda)
 + p(b,B^{\prime})\lw(A,B^{\prime},\P_{\mu+T_{B^{\prime}}}(b,D);\lambda) \nonumber\\
 \vphantom{\Big(}
&-& \P_{\lambda+\mu}(\lw(A,b,B^{\prime};\lambda),D)
 - \rw(\P_\lambda(A,b),B^{\prime},D;\lambda+\mu) \nonumber\\
 \vphantom{\Big(}
&+& p(A,b)\N(b,\j(A,B^{\prime},D;\lambda,\mu+T_b)) \label{a9}\\
\vphantom{\Big(}
&+& p(A,B^{\prime})p(b,B^{\prime})\N(B^{\prime},\j(A,b,D;\lambda,\mu+T_{B^{\prime}})) \nonumber\\
&+& \int_0^\lambda d\nu\ \j(\P_\lambda(A,b),B^{\prime},D;\nu,\lambda+\mu-\nu) \nonumber\\
&+& p(b,B^{\prime}) \int_0^\mu d\nu\ \j(A,B^{\prime},\P_{\mu-\nu}(b,D);\lambda,\nu) \nonumber\\
&+& p(A,B^{\prime})p(b,B^{\prime})\int_0^\mu d\nu\  \P_\nu(B^{\prime},\j(A,b,D;\lambda,\mu-\nu))
\nonumber
\end{eqnarray}
% a10
\begin{eqnarray}
\vphantom{\Big(}
\j(\N(a,A^{\prime}),B,C;\lambda,\mu)
&=& -p(a,B)p(A^{\prime},B)\j(B,\N(a,A^{\prime}),C;\mu,\lambda) \nonumber\\
\vphantom{\Big(}
&-& \P_{\lambda+\mu}(\skewl(\N(a,A^{\prime}),B;\lambda),C) \label{a10}
\end{eqnarray}
% a11
\begin{eqnarray}
\skewn(a,\N(b,B^{\prime}),C)
&=& \skewn(a,b,\N(B^{\prime},C)) + p(a,b)\N(b,\skewn(a,B^{\prime},C)) \label{a11}\\
&-& p(a,b)p(a,B^{\prime})\N(\Big(\int_0^Td\lambda\ b\Big),\lw(B^{\prime},a,C;\lambda)) \nonumber\\
&-& \int_{-T_a-T_b-T_{B^{\prime}}}^0 d\lambda\ \qa(\P_\lambda(a,b),B^{\prime},C) \nonumber\\
&+& p(a,b)\int_0^{T_{B^{\prime}}}d\lambda\ \skewn(\P_\lambda(b,a),B^{\prime},C) \nonumber\\
&-& p(a,b)p(a,B^{\prime}) \N(\Big(\int_{0}^{T_b}d\lambda\ b\Big),\N(\skewl(B^{\prime},a;\lambda),C))
 \nonumber\\
&+& \skewn(a,\Big(\int_0^Td\lambda\ b\Big),\P_\lambda(B^{\prime},C)) \nonumber\\
&+& p(b,B^{\prime})\skewn(a,\Big(\int_0^Td\lambda\ B^{\prime}\Big),\P_\lambda(b,C)) \nonumber\\
&-& p(a,b) \N(\Big(\int_0^Td\lambda\int_0^\lambda d\mu\ \ b\Big),
 \P_\mu(\skewl(a,B^{\prime};\mu-\lambda),C)) \nonumber
\end{eqnarray}
% a12
\begin{eqnarray}
\skewn(\N(a,A^{\prime}),B,C)
&=& \N(a,\skewn(A^{\prime},B,C)) + p(A^{\prime},B)\skewn(a,B,\N(A^{\prime},C)) \nonumber\\
&+& \N(\Big(\int_0^Td\lambda\ a\Big),\lw(A^{\prime},B,C;\lambda)) \label{a12}\\
&+& p(a,A^{\prime})\N(\Big(\int_0^Td\lambda\ A^{\prime}\Big),\lw(a,B,C;\lambda)) \nonumber\\
&-& p(a,A^{\prime})\int_{-T_a-T_{A^{\prime}}-T_B}^0 d\lambda\  \qa(\Big(e^{T\partial_\lambda}A^{\prime}\Big),\P_\lambda(a,B),C) \nonumber\\
&+& p(A^{\prime},B)\skewn(\Big(\int_{-T}^0d\lambda\ \P_\lambda(a,B)\Big),A^{\prime},C) \nonumber\\
&+& p(A^{\prime},B)\skewn(\Big(\int_0^Td\lambda\ a\Big),B,\P_\lambda(A^{\prime},C)) \nonumber\\
&+& p(a,A^{\prime})p(a,B)\skewn(\Big(\int_0^Td\lambda\ A^{\prime}\Big),B,\P_\lambda(a,C)) 
\nonumber\\
&+& p(a,A^{\prime})\N(\Big(\int_{-T}^0d\mu\ \skewl(A^{\prime},
 \Big(\int_{-T}^0d\lambda\ \P_\lambda(a,B)\Big);\mu)\Big),C) \nonumber
\end{eqnarray}
% a13
\begin{eqnarray}
\skewl(a,\N(b,B^{\prime});\lambda)
&=& p(a,b)\N(b,\skewl(a,B^{\prime};\lambda)) + \skewn(\P_\lambda(a,b),B^{\prime},1) \nonumber\\
&-& p(a,b)\int_{-T}^\lambda d\mu\ \skewl(\P_{\mu-\lambda}(b,a),B^{\prime};\mu) \label{a13}
\end{eqnarray}
% a14
\begin{eqnarray}
\vphantom{\Big(}
\skewl(\N(a,A^{\prime}),B;\lambda)
&=& p(A^{\prime},B)\N(\skewl(a,B;\lambda+T_{A^{\prime}}),A^{\prime}) \label{a14}\\
\vphantom{\Big(}
&+& \N(\Big(e^{T\partial_\lambda}a\Big),\skewl(A^{\prime},B;\lambda)) \nonumber\\
\vphantom{\Big(}
&+& p(a,B)p(A^{\prime},B)\lw(B,a,A^{\prime};-\lambda-T) \nonumber\\
\vphantom{\Big(}
&+& p(a,A^{\prime})\skewn(\Big(e^{T\partial_\lambda}A^{\prime}\Big),\P_\lambda(a,B),1) \nonumber\\
\vphantom{\Big(}
&-& p(a,A^{\prime})p(a,B)\int_{-T}^\lambda d\mu\ 
 \skewl(\Big(e^{T\partial_\lambda}A^{\prime}\Big),\P_{-\lambda-T}(B,a);\mu) \nonumber\\
 \vphantom{\Big(}
&+& p(a,A^{\prime})\int_{-T}^\lambda d\mu\ \P_\mu(A^{\prime},\skewl(a,B;\lambda-\mu)) \nonumber
\end{eqnarray}

% - corollary 1: two sided ideal

\begin{corollary}\label{basic}
For every $\Delta_1,\Delta_2,\Delta_3\in\Gamma$ there exists $\Delta^\prime\in\Gamma$ 
such that the following inclusions hold
\begin{eqnarray}
\!\!\!\!\!\!\!\!\!
\vphantom{\Big(}
\N(\T_{\Delta_1}(R),\M_{\Delta_2}(R)) &\subset& \M_{\Delta_1+\Delta_2}(R)\ , \label{b1}\\
\!\!\!\!\!\!\!\!\!
\vphantom{\Big(}
\N(\M_{\Delta_1}(R),\T_{\Delta_2}(R)) &\subset& \M_{\Delta_1+\Delta_2}(R)\ , \label{b2}\\
\!\!\!\!\!\!\!\!\!
\vphantom{\Big(}
\P_\lambda(\T_{\Delta_1}(R),\M_{\Delta_2}(R)) 
&\subset& \C[\lambda]\otimes\M_{\Delta^\prime}(R)\ ,\ 
\Delta^\prime<\Delta_1\!+\!\Delta_2\ ,\label{b3}\\
\!\!\!\!\!\!\!\!\!
\vphantom{\Big(}
\P_\lambda(\M_{\Delta_1}(R),\T_{\Delta_2}(R)) 
&\subset& \C[\lambda]\otimes\M_{\Delta^\prime}(R)\ ,\ 
\Delta^\prime<\Delta_1\!+\!\Delta_2\ , \label{b4}%\\
\end{eqnarray}
\begin{eqnarray}%
\!\!\!\!\!\!\!\!\!
\vphantom{\Big(}
\qa(\T_{\Delta_1}(R),\T_{\Delta_2}(R),\T_{\Delta_3}(R)) 
&\subset& \M_{\Delta_1+\Delta_2+\Delta_3}(R)\ , \label{b5}\\
\!\!\!\!\!\!\!\!\!
\vphantom{\Big(}
\lw(\T_{\Delta_1}\!(R),\!\T_{\Delta_2}\!(R),\!\T_{\Delta_3}\!(R);\!\lambda) 
&\subset& \C[\lambda]\!\otimes\!\M_{\Delta^\prime},\
\Delta^\prime<\Delta_1\!\!+\!\!\Delta_2\!\!+\!\!\Delta_3\ , \label{b6}\\
\!\!\!\!\!\!\!\!\!
\vphantom{\Big(}
\rw(\T_{\Delta_1}\!(R),\!\T_{\Delta_2}\!(R),\!\T_{\Delta_3}\!(R);\!\lambda) 
&\subset& \C[\lambda]\!\otimes\!\M_{\Delta^\prime}(R),\ 
\Delta^\prime\!<\!\Delta_1\!\!+\!\!\Delta_2\!\!+\!\!\Delta_3, \label{b7}\\
\!\!\!\!\!\!\!\!\!
\vphantom{\Big(}
\j(\T_{\Delta_1}\!(R),\!\T_{\Delta_2}\!(R),\!\T_{\Delta_3}\!(R);\!\lambda,\!\mu) 
&\subset& \C[\lambda,\mu]\!\otimes\!\M_{\Delta^\prime},\ 
\Delta^\prime<\Delta_1\!\!+\!\!\Delta_2\!\!+\!\!\Delta_3, \label{b8}\\
\!\!\!\!\!\!\!\!\!
\vphantom{\Big(}
\skewn(\T_{\Delta_1}(R),\T_{\Delta_2}(R),\T_{\Delta_3}(R)) 
&\subset& \M_{\Delta_1+\Delta_2+\Delta_3}(R)\ , \label{b9}\\
\!\!\!\!\!\!\!\!\!
\vphantom{\Big(}
\skewl(\T_{\Delta_1}(R),\T_{\Delta_2}(R);\lambda) 
&\subset& \C[\lambda]\otimes\M_{\Delta^\prime}\ ,\ 
\Delta^\prime<\Delta_1\!+\!\Delta_2\ .\label{b10}
\end{eqnarray}
In particular, $\M(R)$ is a two sided ideal 
with respect to both the normally ordered product 
$\N:\ \T(R)\otimes\T(R)\rightarrow\T(R)$ and the $\lambda$--bracket
$\P_\lambda:\ \T(R)\otimes\T(R)\rightarrow\T(R)$.
\end{corollary}

% proof of the corollary

\begin{proof}
We will prove, by induction on $\Delta\in\Gamma$, that there exists $\Delta^\prime<\Delta$
such that the following conditions hold, for 
$A\in\T_{\Delta_A}(R),\ B\in\T_{\Delta_B}(R),\ C\in\T_{\Delta_C}(R),\ D\in\T_{\Delta_D}(R)$ 
and $E\in\M_{\Delta_E}(R)$:
\begin{eqnarray}
\vphantom{\Big(}
\qa(A,B,C) &\in& \M_\Delta(R)\ ,
\quad \text{ if } \Delta_A+\Delta_B+\Delta_C\leq\Delta\ , \label{c5}\\
\vphantom{\Big(}
\lw(A,B,C;\lambda) &\in& \C[\lambda]\M_{\Delta^\prime}(R)\ ,
\quad \text{ if } \Delta_A+\Delta_B+\Delta_C\leq\Delta\ , \label{c6}\\
\vphantom{\Big(}
\rw(A,B,C;\lambda) &\in& \C[\lambda]\M_{\Delta^\prime}(R)\ ,
\quad \text{ if } \Delta_A+\Delta_B+\Delta_C\leq\Delta\ , \label{c7}\\
\vphantom{\Big(}
\j(A,B,C;\lambda,\mu) &\in& \C[\lambda,\mu]\M_{\Delta^\prime}(R)\ ,
\quad \text{ if } \Delta_A+\Delta_B+\Delta_C\leq\Delta\ , \label{c8}\\
\vphantom{\Big(}
\N(A,E) &\in& \M_\Delta(R)\ ,
\quad \text{ if } \Delta_A+\Delta_E\leq\Delta\ ,\label{c1}\\
\vphantom{\Big(}
\N(E,D) &\in& \M_\Delta(R)\ ,
\quad \text{ if } \Delta_D+\Delta_E\leq\Delta\ ,\label{c2}\\
\vphantom{\Big(}
\P_\lambda(A,E) &\in& \C[\lambda]\otimes\M_{\Delta^\prime}(R)\ ,
\quad \text{ if } \Delta_A+\Delta_E\leq\Delta\ ,\label{c3}\\
\vphantom{\Big(}
\P_\lambda(E,D) &\in& \C[\lambda]\otimes\M_{\Delta^\prime}(R)\ ,
\quad \text{ if } \Delta_D+\Delta_E\leq\Delta\ ,\label{c4}\\
\vphantom{\Big(}
\skewn(A,B,C) &\in& \M_\Delta(R)\ ,
\quad \text{ if } \Delta_A+\Delta_B+\Delta_C\leq\Delta\ , \label{c9}\\
\vphantom{\Big(}
\skewl(A,B;\lambda) &\in& \C[\lambda]\otimes\M_{\Delta^\prime}(R)\ ,
\quad \text{ if } \Delta_A+\Delta_B\leq\Delta\ , \label{c10}\ .
\end{eqnarray}
For $\Delta=0$ there is nothing to prove. Let then $\Delta>0$ and assume, by induction, 
that the above conditions hold for all $\bar{\Delta}<\Delta$.

% c5
{\itshape proof of (\ref{c5})}\\
For $A\in\C$ condition (\ref{c5}) is trivially satisfied, and for $A\in R$ it holds thanks to equation (\ref{3}).
Let then $A=a\otimes A^\prime$, with $a\in R$ and $A^\prime\in R\oplus R^{\otimes2}\oplus\cdots$.
In this case $\qa(A,B,C)$ is given by equation (\ref{a4}), and every term in the right hand side belongs
to $\M_\Delta(R)$ by the inductive assumption.

% c6
{\itshape proof of (\ref{c6})}\\
For $A\in\C$ there is nothing to prove. We will consider separately two cases:
\begin{enumerate}
\item $A=a\in R$,
\item $A=a\otimes A^\prime$, with $a\in R$ and $A^\prime\in R\oplus R^{\otimes2}\oplus\cdots$.
\end{enumerate}

\noindent {\itshape Case 1.}
For $B\in\C$ or $B\in R$ condition (\ref{c6}) is trivially satisfied, 
thanks to equation (\ref{6}).
Let then $B=b\otimes B^\prime$, with $b\in R$ and $B^\prime\in R\oplus R^{\otimes2}\oplus\cdots$.
In this case $\lw(A,B,C;\lambda)$ is given by equation (\ref{a5}), and, by the inductive assumption,
every term of the right hand side lies in $\M_{\Delta^\prime}(R)$, for some $\Delta^\prime<\Delta$.

\noindent {\itshape Case 2.}
In this case $\lw(A,B,C;\lambda)$ is given by equation (\ref{a6}), and again, by induction,
every term of the right hand side lies in $\M_{\Delta^\prime}(R)$, for some $\Delta^\prime<\Delta$.

% c7
{\itshape proof of (\ref{c7})}\\
For $A\in\C$ or $A\in R$, condition (\ref{c7}) holds trivially, thanks to equation (\ref{7}).
Suppose then 
$A=a\otimes A^\prime$, with $a\in R$ and $A^\prime\in R\oplus R^{\otimes2}\oplus\cdots$.
In this case $\rw(A,B,C;\lambda)$ is given by equation (\ref{a7}), and every term in the right hand side
lies in $\M_{\Delta^\prime}(R)$, for some $\Delta^\prime<\Delta$.

% c8
{\itshape proof of (\ref{c8})}\\
If either $A\in\C$, or $B\in\C$, or $C\in\C$, $\j(A,B,C;\lambda,\mu)=0$, so that (\ref{c8}) holds.
Moreover, if $A,B,C\in R$, condition (\ref{c8}) holds by definition of non--linear Lie conformal algebra.
We will consider separately the following three cases:
\begin{enumerate}
\item $A=a\in R,\ B=b\in R,\ C=c\otimes D$, 
with $c\in R$ and $D\in R\oplus R^{\otimes2}\oplus\cdots$,
\item $B=b\otimes B^\prime$, 
with $b\in R$ and $A,B^\prime,C\in R\oplus R^{\otimes2}\oplus\cdots$,
\item $A=a\otimes A^\prime$, 
with $a\in R$ and $A^\prime,B,C\in R\oplus R^{\otimes2}\oplus\cdots$.
\end{enumerate}
In the first case $\j(A,B,C;\lambda,\mu)$ is expressed by equation (\ref{a8}),
and, by the inductive assumption,
every term in the right hand side lies in $\M_{\Delta^\prime}(R)$,
for some $\Delta^\prime<\Delta$.
In the second case $\j(A,B,C;\lambda,\mu)$ is given by equation (\ref{a9}), 
and it lies by induction in $\M_{\Delta^\prime}(R)$, for some $\Delta^\prime<\Delta$.
Finally, the third case reduces to the second case, thanks to equation (\ref{a10}) and 
the inductive assumption.

% c1
{\itshape proof of (\ref{c1})}\\
For $A\in\C$ condition (\ref{c1}) is obvious, and for $A\in R$ it holds by definition of $\M_\Delta(R)$.
Let then $A=a\otimes B$, with $a\in R$ and $B\in R\oplus R^{\otimes2}\oplus\cdots$.
By (\ref{3}) we have
\begin{eqnarray*}
\N(A,E) &=& a\otimes\N(B,C)
 +\N(\Big(\int_0^Td\lambda\ a\Big), \P_\lambda(B,E)) \\
&+& p(a,b)\N(\Big(\int_0^Td\lambda\ B\Big)\P_\lambda(a,E)\ ,
\end{eqnarray*}
and each term in the right hand side is in $\M_\Delta(R)$ by the inductive assumption.

% c2
{\itshape proof of (\ref{c2})}\\
We will consider separately the following two cases:
\begin{enumerate}
\item $E=a\otimes F$, with $a\in R$ and $F\in\M_{\Delta_F}$, where $\Delta_a+\Delta_F=\Delta_E$,
\item $E=\skewn(a,b,C)$, with $a,b\in R,\ C\in\T(R)$, 
and $\Delta_a+\Delta_b+\Delta_C=\Delta_E$.
\end{enumerate}
In the first case we have, by (\ref{3}),
\begin{eqnarray*}
\N(E,D) &=& a\otimes\N(F,D))
 +\N(\Big(\int_0^Td\lambda\ a\Big), \P_\lambda(F,D)) \\
&+& p(a,F)\N(\Big(\int_0^Td\lambda\ F\Big)\P_\lambda(a,D)\ ,
\end{eqnarray*}
and each term in the right hand side is in $\M_\Delta(R)$ by the inductive assumption.
In the second case $\N(E,D)$ is expressed by equation (\ref{a1}),
and again each term in the right hand side lies in $\M_\Delta(R)$.

% c3
{\itshape proof of (\ref{c3})}\\
For $A\in\C$ condition (\ref{c3}) is obvious. We will consider separately two cases:
\begin{enumerate}
\item $A=a\in R$,
\item $A=a\otimes B$, with $a\in R$ and $B\in R\oplus R^{\otimes2}\oplus\cdots$.
\end{enumerate}
Consider the second case first. We have, by (\ref{7}),
\begin{eqnarray*}
\P_\lambda(A,E) 
&=& \N(\Big(e^{T\partial_\lambda}a\Big),\P_\lambda(B,E)) 
+ p(a,B)\N(\Big(e^{T\partial_\lambda}B\Big),\P_\lambda(a,E)) \\
&+& p(a,B)\int_0^\lambda d\mu\  \P_\mu(B,\P_{\lambda-\mu}(a,E)) \ ,
\end{eqnarray*}
and, by induction, each term in the right hand side lies in $\M_{\Delta^\prime}(R)$, 
for some $\Delta^\prime<\Delta$.
Suppose then $A=a\in R$. As before, there are two different situations to consider:
\renewcommand{\theenumi}{\alph{enumi}}
\renewcommand{\labelenumi}{2\theenumi}
\begin{enumerate}
\item $E=b\otimes F$, with $b\in R$, $F\in\M_{\Delta_F}$ and $\Delta_b+\Delta_F=\Delta_E$,
\item $E=\skewn(b,c,D)$, with $b,c\in R,\ D\in\T(R)$, 
and $\Delta_b+\Delta_c+\Delta_D=\Delta_E$.
\end{enumerate}
\renewcommand{\theenumi}{\arabic{enumi}}
\renewcommand{\labelenumi}{\theenumi.}
In the first case we have, by (\ref{6}),
\begin{eqnarray*}
\P_\lambda(a,b\otimes F) 
&=& \N(\P_\lambda(a,b),F)+p(a,b)\N(b,\P_\lambda(a,F)) \\
&+& \int_0^\lambda d\mu\  \P_\mu(\P_{\lambda}(a,b),F) \ ,
\end{eqnarray*}
and each term in the right hand side lies by induction in $\M_{\Delta^\prime}(R)$,
for some $\Delta^\prime<\Delta$.
Finally, if $E=\skewn(b,c,D)$, $\P_\lambda(a,E)$ is expressed by equation (\ref{a2}),
and again each term in the right hand side lies in $\M_{\Delta^\prime}(R)$,
for some $\Delta^\prime<\Delta$.

% c4
{\itshape proof of (\ref{c4})}\\
Consider separately two cases:
\begin{enumerate}
\item $E=a\otimes F$, with $a\in R$, $F\in\M_{\Delta_F}$, and $\Delta_a+\Delta_F=\Delta_E$,
\item $E=\skewn(a,b,C)$, with $a,b\in R,\ C\in\T(R)$, 
and $\Delta_a+\Delta_b+\Delta_C=\Delta_E$.
\end{enumerate}
In the first case we have, by (\ref{7}),
\begin{eqnarray*}
\P_\lambda(E,D) 
&=& \N(\Big(e^{T\partial_\lambda}a\Big),\P_\lambda(F,D)) 
+ p(a,B)\N(\Big(e^{T\partial_\lambda}F\Big),\P_\lambda(a,D)) \\
&+& p(a,B)\int_0^\lambda d\mu\  \P_\mu(F,\P_{\lambda-\mu}(a,D))\ , 
\end{eqnarray*}
and, by the inductive assumption,
each term in the right hand side is in $\M_{\Delta^\prime}(R)$,
form some $\Delta^\prime<\Delta$.
In the second case $\P_\lambda(E,D)$ is given by equation (\ref{a3}).
All the terms of the right hand side are in $\M_{\Delta^\prime}(R)$,
for some $\Delta^\prime<\Delta$, thanks to the inductive assumption
and condition (\ref{c8}).

% c9
{\itshape proof of (\ref{c9})}\\
For $A\in\C$ or $B\in\C$, condition (\ref{c9}) holds trivially.
Moreover, for $A,B\in R$, (\ref{c9}) holds by definition of $\M_\Delta(R)$.
%a
Suppose first $A=a\in R,\ B=b\otimes B^{\prime}$, with $b\in R$ 
and $B^\prime\in R\oplus R^{\otimes2}\oplus\cdots$.
In this case $\skewn(A,B,C)$ is given by equation (\ref{a11}), and it belongs to
$\M_\Delta(R)$ by induction.
%b
We are left to consider the case
$A=a\otimes A^{\prime}$, with $a\in R$ 
and $A^\prime,B\in R\oplus R^{\otimes2}\oplus\cdots$.
In this case $\skewn(A,B,C)$ is given by equation (\ref{a12}). Every term in the right hand side
lies in $\M_\Delta(R)$ thanks to the inductive assumption and the above result.

% c10
{\itshape proof of (\ref{c10})}\\
If either $A\in\C$, or $B\in\C$, or $A,B\in R$, condition (\ref{c10}) is trivial.
Suppose $A=a\in R$ and  $B=b\otimes B^{\prime}$, with $b\in R$ 
and $B^\prime\in R\oplus R^{\otimes2}\oplus\cdots$.
In this case $\skewl(A,B;\lambda)$ is given by equation (\ref{a13}),
and it lies by induction in $\M_{\Delta^\prime}(R)$,
for some $\Delta^\prime<\Delta$.
Finally, consider the case $A=a\otimes A^{\prime}$, with $a\in R$ 
and $A^\prime,B\in R\oplus R^{\otimes2}\oplus\cdots$.
In this case $\skewl(A,B;\lambda)$ is given by equation (\ref{a14}),
and again it lies in $\M_{\Delta^\prime}(R)$, for some $\Delta^\prime<\Delta$, 
by the inductive assumption and by condition (\ref{c6}).
\end{proof}

% relationf for the Lie bracket

\begin{corollary}\label{lie}
The map $\P:\ \T(R)\otimes\T(R)\rightarrow\T(R)$ defined in Remark \ref{liebrack} satisfies skewsymmetry:
$$
\P(A,B)+p(A,B)\P(B,A)\ \in\ \M_{\Delta^\prime}(R)\ ,
$$
for every $A,B\in\T(R)$ and for some $Delta^\prime<\Delta(A)+\Delta(B)$,
and Jacobi identity,
$$
\P(A,\P(B,C))-p(A,B)\P(B,\P(A,C))-\P(\P(A,B),C)\ \in\ \M_{\Delta^\prime}(R)\ ,
$$
for every $A,B,C\in\T(R)$ and for some  $\Delta^\prime<\Delta(A)+\Delta(B)+\Delta(C)$.
\end{corollary}
\begin{proof}
The corollary follows immediately from Corollary \ref{basic} and the following identities
\begin{eqnarray*}
\P(A,B) &=& -p(A,B)\P(B,A) + \int_{-T}^0d\lambda\ \skewl(A,B;\lambda)\ ,\\
\P(A,\P(B,C)) &=& p(A,B)\P(B,\P(A,C)) + \P(\P(A,B),C) \\
&+& \int_{-T}^0d\lambda\int_{-\lambda-T}^0d\mu\ 
\j(A,B,C;\lambda,\mu)\ .
\end{eqnarray*}
\end{proof}

%%%%%%%% SECTION 5 %%%%%%%%%%%%%%%%%%%%%%%%%%%%%%%%

\section{Proof of Theorem \ref{main}}
\setcounter{equation}{0}

% II part

In this section we will use Corollary \ref{basic} to prove Theorem \ref{main}.
Notice that by (\ref{b1})--(\ref{b4}), the products $\N:\ \T(R)\otimes\T(R)\rightarrow\T(R)$
and $\P_\lambda:\ \T(R)\otimes\T(R)\rightarrow\C[\lambda]\otimes\T(R)$
induce products on the quotient space $U(R)=\T(R)/\M(R)$,
which we denote by $:\ \ :$ and $[\ _\lambda\ ]$ respectively.
Moreover, by (\ref{b5})--(\ref{b10}), the space $U(R)$, with normally ordered product $:\ \ :$ and
$\lambda$--bracket $[\ _\lambda\ ]$, satisfies all the axioms of vertex algebra.
This proves the second part of Theorem \ref{main}.

% I part

We are left to prove the first part of Theorem \ref{main}, which provides a PBW basis for $U(R)$,
and therefore guarantees that $U(R)\neq0$.
Let us denote by $\tB$ the collection of 1 and all ordered monomials in $\T(R)$:
$$
\tB\ =\
\bigg\{
a_{i_1}\otimes\dots\otimes a_{i_n}
\left|\begin{array}{c}
i_1\leq\dots\leq i_n,\  n\in\Z_+\ ,\\
i_k<i_{k+1} \text{ if } p(a_{i_k})=\bar{1}
\end{array}\right\}
$$
We also denote by $\tB[\Delta]$ the collection of ordered monomials of degree $\Delta$,
namely $\tB[\Delta]=\tB\cap\T(R)[\Delta]$,
and by $\tB_\Delta$ the corresponding filtration, namely
$\tB_\Delta=\tB\cap\T_\Delta(R)$.
By definition $\B=\pi(\tB)$.

% easy part

\begin{lemma}\label{l1}
Any element $E\in\T_\Delta(R)$ can be decomposed as 
\begin{equation}\label{dec}
E=P+M\ ,
\end{equation}
where $P\in\Span_\C\tB_\Delta$ and $M\in\M_\Delta(R)$.
Equivalently
$$
\T_\Delta(R)/\M_\Delta(R)\ =\ \Span_\C\pi(\tB_\Delta)\ ,\quad\forall\Delta\in\Gamma\ ,
$$
and therefore $U(R)=\Span_\C\B$.
\end{lemma}
\begin{proof}
It suffices to prove (\ref{dec}) for monomials, namely for
$E=a_{j_1}\otimes\dots\otimes a_{j_n}\in\T(R)[\Delta]$,
where $\Delta\in\Gamma$ and $a_{j_k}\in\A$.
Let us define the number of {\itshape inversions} of $E$ as
$$
d(E)=\#\left\{(p,q)\ \ \left|\ \ \begin{array}{c}
1\leq p<q\leq n \text{ and } \\
\text{ either } a_{j_p} \text{ is even and } j_p>j_q\\
\text{ or } a_{j_p} \text{ is odd and } j_p\geq j_q
\end{array}\right.\right\}\ .
$$
Let $d=d(E)$.
We will prove that $E$ decomposes as in (\ref{dec}) by induction 
on the pair $(\Delta,d)$, ordered lexicographically.
If $d=0$, we have 
$E\in\tB[\Delta]$, so there is nothing to prove.
Suppose that $d\geq1$ and let $p\in\{1,\dots,n-1\}$ 
be such that $j_p>j_{p+1}$, or $j_p=j_{p+1}$ for $a_{j_p}$ odd.
In the first case we have, by definition of $\M_\Delta(R)$,
\begin{eqnarray*}
E &\equiv& p(a_{j_p},a_{j_{p+1}})a_{j_1}\otimes\dots
 \otimes a_{j_{p+1}}\otimes a_{j_{p}}\otimes\dots\otimes a_{j_n} \\
&+& a_{j_1}\otimes\dots\otimes
 \N(\Big(\int_{-T}^0d\lambda[a_{j_{p}}\ _\lambda\ a_{j_{p+1}}]\Big),
 \dots\otimes a_{j_n})\quad \mod\M_\Delta(R) \ .
\end{eqnarray*}
The first term in the right hand side has degree $\Delta$ and $d-1$ inversions,
while, by the grading conditions (\ref{8}), the second term lies in $\T_{\Delta^\prime}(R)$,
for some $\Delta^\prime<\Delta$.
Therefore, by the inductive assumption, both terms admit a decomposition (\ref{dec}).
Consider now the second case, namely $a_{j_p}$ is odd and $j_p=j_{p+1}$..
By definition of $\M_\Delta(R)$ we have
$$
E \equiv \frac{1}{2}a_{j_1}\otimes\dots\otimes
 \N(\Big(\int_{-T}^0d\lambda[a_{j_{p}}\ _\lambda\ a_{j_{p+1}}]\Big),
 \dots\otimes a_{j_n})\quad \mod\M_\Delta(R) \ ,
$$
and again, by the inductive assumption, the right hand side admits a decomposition (\ref{dec}).
\end{proof}

% key lemma

\begin{lemma}\label{l2}
Let $\tU$ be a vector space with basis $\tB$:
$$
\tU=\bigoplus_{A\in\tB}\C A \ .
$$
There exists a unique linear map
$$
\sigma\ :\ \T(R)\ \longrightarrow\ \tU
$$
such that
\begin{enumerate}
\item $\sigma(A)=A,\ \forall A\in\tB$,
\item $\M(R)\subset\ker\sigma$.
\end{enumerate}
\end{lemma}
\begin{proof}
We want to prove that there is a unique collection of linear maps
$\sigma_\Delta\ :\ \T_\Delta(R) \rightarrow \tU$, for $\Delta\in\Gamma$,
such that
\renewcommand{\labelenumi}{(\theenumi)}
\begin{enumerate}
\item $\sigma_0(1)=1\ ,\quad 
\sigma_\Delta\big|_{\T_{\Delta^\prime}(R)}=\sigma_{\Delta^\prime},\ \text{ if } \Delta^\prime\leq\Delta$,
\item $\sigma_\Delta(A)=A,\ \text{ if } A\in\tB[\Delta]$,
\item $\M_\Delta(R)\subset\ker\sigma_\Delta$.
\end{enumerate}
\renewcommand{\labelenumi}{\theenumi.}
This obviously proves the lemma. Indeed for any such sequence we can define the map 
$\sigma:\T(R)\rightarrow\tU$ by
\begin{equation}\label{s}
\sigma\big|_{\T_\Delta(R)}=\sigma_\Delta,\ \forall\Delta\in\Gamma\ ,
\end{equation}
and conversely, given a map $\sigma:\T(R)\rightarrow\tU$ satisfying the assumptions
of the lemma, we can define such a sequence of maps $\sigma_\Delta:\ \T_\Delta(R)\rightarrow\tU$
by equation (\ref{s}).

The condition $\sigma_0(1)=1$ defines completely $\sigma_0$.
Notice that, since $\M_0(R)=0$, $\sigma_0$ satisfies all the required conditions.
Let then $\Delta>0$, and suppose by induction that $\sigma_{\Delta^\prime}$
is uniquely defined and it satisfies all conditions (1)--(3)
for every $\Delta^\prime<\Delta$.
\smallskip

{\itshape Uniqueness of $\sigma_\Delta$}
\smallskip

\noindent Given a monomial of degree $\Delta$,
$E=a_{j_1}\otimes\dots\otimes a_{j_n}\in\T(R)[\Delta]$, 
we will show that $\sigma_\Delta(E)$ is uniquely defined by induction on 
the number of inversions $d$
of $E$, defined above.
For $d=0$ we have $E\in\tB_\Delta$, so it must be $\sigma_\Delta(E)=E$ 
by condition (2).
Let then $d\geq1$ and let $(j_p,j_{p+1})$ be the ``most left inversion'',
namely $p\in\{1,\dots,n\}$ is the smallest integer such that either $j_p>j_{p+1}$
or $p(a_{j_p})=\bar{1}$ and $j_p=j_{p+1}$.
By condition (3) and the definition of $\M_\Delta(R)$ we then have, if $j_p>j_{p+1}$,
\begin{eqnarray}
\sigma_\Delta(E) &=& p(a_{j_p},a_{j_{p+1}})
 \sigma_\Delta\Big(a_{j_1}\otimes\dots
 \otimes a_{j_{p+1}}\otimes
 a_{j_{p}}\otimes\dots\otimes a_{j_n}\Big) \label{un1}\\
&+& \sigma_\Delta\Big(a_{j_1}\otimes\dots\otimes
\N(\P(a_{j_p},a_{j_{p+1}}),
%\Big(\int_{-T}^0d\lambda [a_{j_p}\ _\lambda\ a_{j_{p+1}}]\Big),
\dots\otimes a_{j_n})\Big) \ , \nonumber
\end{eqnarray}
while, if $p(a_{j_p})=\bar{1}$ and $j_p=j_{p+1}$,
\begin{equation}\label{un2}
\sigma_\Delta(E)\ 
=\ \frac{1}{2}\sigma_\Delta\Big(a_{j_1}\otimes\dots\otimes
\N(\P(a_{j_{p}},a_{j_{p+1}}),
%\Big(\int_{-T}^0d\lambda [a_{j_p}\ _\lambda\ a_{j_{p+1}}]\Big),
\dots\otimes a_{j_n})\Big) \ ,
\end{equation}
Notice that  $(a_{j_1}\otimes\dots \otimes a_{j_{p+1}}\otimes
a_{j_{p}}\otimes\dots \otimes a_{j_n})$ has degree $\Delta$ and $d-1$ inversions, 
so the first term in the right hand side of (\ref{un1}) is uniquely defined by the inductive
assumption.
Moreover, by the grading conditions (\ref{8}),
$a_{j_1}\otimes\dots\otimes\N(\P(a_{j_{p}},a_{j_{p+1}}),
%\Big(\int_{-T}^0d\lambda [a_{j_p}\ _\lambda\ a_{j_{p+1}}]\Big),
\dots\otimes a_{j_n})$ lies in $\T_{\Delta^\prime}(R)$,
for some $\Delta^\prime<\Delta$, so the second term in the right hand side 
of (\ref{un1}) and the right hand side of (\ref{un2}) are uniquely defined 
by condition (1) and the inductive assumption.
\smallskip

{\itshape Existence of $\sigma_\Delta$}
\smallskip

\noindent The above prescription defines uniquely a linear map
$\sigma_\Delta:\ \T_\Delta(R)\rightarrow\tU$, which by construction
satisfies conditions (1) and (2).
We are left to show that condition (3) holds.
By definition of $\M_\Delta(R)$, it suffices to prove that, for every 
monomial of degree $\Delta$, $E=a_{j_1}\otimes\dots\otimes a_{j_n}\in\T(R)[\Delta]$,
and for every $q=1,\dots,n$, we have
\begin{equation}\label{jacobson1}
\sigma_\Delta\Big(a_{j_1}\otimes\cdots a_{j_{q-1}}
\otimes \skewn(a_{j_{q}},a_{j_{q+1}},
a_{j_{q+2}}\otimes\dots\otimes a_{j_n})\Big)=0 \ .
\end{equation}
By skewsymmetry of the $\lambda$--bracket we can assume, without loss of generality,
that $j_q\geq j_{q+1}$. Moreover, if $p(a_{j_q})=\bar{0}$, we have
$\int_{-T}^0d\lambda[a_{j_q}\ _\lambda\ a_{j_{q}}]=0$, so that when $j_q=j_{q+1}$
equation (\ref{jacobson1}) is obvious.
In other words, we can assume that $(j_q,j_{q+1})$ is an "inversion" of $E$.
In particular $d=d(E)\geq1$.
We will prove condition (\ref{jacobson1}) by induction on $d$.
Let $(j_p,j_{p+1})$ be the ``most left inversion'' of $E$.
For $p=q$ equation (\ref{jacobson1}) holds by construction.
We will consider separately the cases $p\leq q-2$ and $p=q-1$.
\smallskip

\noindent {\itshape Case 1.} Assume $p\leq q-2$. For simplicity we rewrite
$$
E=A\otimes c\otimes b\otimes D\otimes f\otimes e\otimes H \ ,
$$
where $A=a_{j_1}\otimes\dots\otimes a_{j_{p-1}}$,
$c=a_{j_{p}}$, $b=a_{j_{p+1}}$,
$D=a_{j_{p+2}}\otimes\dots\otimes a_{j_{q-1}}$,
$f=a_{j_{q}}$, $e=a_{j_{q+1}}$,
$H=a_{j_{q+2}}\otimes\dots\otimes a_{j_n}$.
The left hand side of equation (\ref{jacobson1}) then takes the form
\begin{eqnarray}
&\sigma_\Delta
\Big(A\otimes c\otimes b\otimes 
D\otimes f\otimes e\otimes H\Big)
-p(e,f) \sigma_\Delta\Big(A\otimes c\otimes b\otimes 
D\otimes e\otimes f\otimes H\Big)& \nonumber\\
&-\sigma_\Delta\Big(A\otimes c\otimes b\otimes D\otimes
\N(\P(e,f),
%\Big(\int_{-T}^0d\lambda[e\ _\lambda\ f]\Big),
H)\Big) \ .& \label{jacobson2}
\end{eqnarray}
By definition of $\sigma_\Delta$, the first term of (\ref{jacobson2})
can be written as
\begin{eqnarray}
&p(b,c)\sigma_\Delta(A\otimes b\otimes c\otimes 
D\otimes f\otimes e\otimes H)& \label{jacobson3}\\
&+\sigma_\Delta(A\otimes\N(\P(c,b),
%\Big(\int_{-T}^0d\lambda[c\ _\lambda\ b]\Big),
D\otimes f\otimes e\otimes H)) \ .& \nonumber
\end{eqnarray}
If $e\neq f$, $A\otimes c\otimes b\otimes D\otimes e\otimes f\otimes H$ 
has degree $\Delta$ and $d-1$ inversions, so that
we can use the inductive assumption to rewrite the second term 
of (\ref{jacobson2}) as
\begin{eqnarray}
&-p(e,f)p(b,c)\sigma_\Delta(A\otimes b\otimes c\otimes 
D\otimes e\otimes f\otimes H)& \label{jacobson4} \\
&-p(e,f)\sigma_\Delta(A\otimes \N(\P(c,b),
%\Big(\int_{-T}^0d\lambda[c\ _\lambda\ b]\Big),
D\otimes e\otimes f\otimes H)) \ .& \nonumber
\end{eqnarray}
By the grading conditions (\ref{8}),
$A\otimes c\otimes b\otimes D\otimes\N(\P(f,e),
%\Big(\int_{-T}^0d\lambda[f\ _\lambda\ e]\Big),
H)
\in\T_{\Delta^\prime}(R)$, for some $\Delta^\prime<\Delta$.
We can thus use the fact that 
$\sigma_\Delta\big|_{\T_{\Delta^\prime}(R)}=\sigma_{\Delta^\prime}$
and the inductive assumption to rewrite the third term of (\ref{jacobson2}) as
\begin{eqnarray}
&-p(b,c)\sigma_\Delta(
A\otimes b\otimes c\otimes D\otimes\N(\P(f,e),
%\Big(\int_{-T}^0d\lambda[f\ _\lambda\ e]\Big),
H)&
 \label{jacobson5} \\
&-\sigma_\Delta(A\otimes\N(\P(c,b),
%\Big(\int_{-T}^0d\lambda[c\ _\lambda\ b]\Big),
D\otimes\N(\P(f,e),
%\Big(\int_{-T}^0d\lambda[f\ _\lambda\ e]\Big),
H))) \ .& \nonumber
\end{eqnarray}
Combining (\ref{jacobson3}), (\ref{jacobson4}) and (\ref{jacobson5})
we can rewrite (\ref{jacobson2}) as
\begin{eqnarray}
&p(b,c)\sigma_\Delta(A\otimes b\otimes c\otimes D\otimes\skewn(f,e,H))& \label{jacobson6}\\
&+\sigma_\Delta(A\otimes \N(\P(c,b),
%\Big(\int_{-T}^0d\lambda[c\ _\lambda\ b]\Big),
D\otimes\skewn(f,e,H)))\ .& \nonumber
\end{eqnarray}
The first term of (\ref{jacobson6}) appears only for $b\neq c$, and in this case it is zero 
by the inductive assumption, since $A\otimes b\otimes c\otimes D\otimes f\otimes e\otimes H$ 
has degree $\Delta$ and $d-1$ inversions.
Consider the second term of (\ref{jacobson6}).
By the grading conditions (\ref{8}) and by Corollary \ref{basic}, the argument of $\sigma_\Delta$
lies in $\M_{\Delta^\prime}(R)$, for some $\Delta^\prime<\Delta$.
It follows by induction that also the second term of (\ref{jacobson6})
is zero.
We thus proved, as we wanted, that (\ref{jacobson2}) is zero.
\smallskip

\noindent {\itshape Case 2.} We are left to consider the case $p=q-1$. For simplicity we write
$$
E=A\otimes c\otimes b\otimes a\otimes D\ ,
$$
where $A=a_{j_1}\otimes\dots\otimes a_{j_{p-1}}$,
$c=a_{j_{p}}$, $b=a_{j_{p+1}}$, $a=a_{j_{p+2}}$,
$D=a_{j_{p+3}}\otimes\dots\otimes a_{j_n}$.
The left hand side of equation (\ref{jacobson1}) then takes the form
\begin{eqnarray}
&\sigma_\Delta(A\otimes c\otimes b\otimes a\otimes D)
-p(a,b)\sigma_\Delta(A\otimes c\otimes a\otimes b\otimes D)& \nonumber\\
&-\sigma_\Delta(A\otimes c\otimes\N(\P(b,a),
%\Big(\int_{-T}^0d\lambda[b\ _\lambda\ a]\Big),
D)) \ .& \label{jacobson7}
\end{eqnarray}
After some manipulations based on the inductive assumption, similar to the ones
used above, we can rewrite (\ref{jacobson7}) as
\begin{eqnarray}
&\sigma_\Delta(A\otimes\Big\{
\N(\P(c,b),
%\Big(\!\int_{-T}^0d\lambda[c\ _\lambda\ b]\!\Big),
a\otimes D))
-p(a,b)p(a,c)a\otimes\N(\P(c,b),
%\Big(\!\int_{-T}^0d\lambda[c\ _\lambda\ b]\Big)\!,
D)\Big\})&\label{jacobson8}\\
&+\sigma_\Delta(A\otimes\Big\{
p(b,c)b\otimes\N(\P(c,a),
%\Big(\!\int_{-T}^0d\lambda[c\ _\lambda\ a]\!\Big)\!,
D))
-p(a,b)\N(\P(c,a),
%\Big(\int_{-T}^0d\lambda[c\ _\lambda\ a]\!\Big)\!,
b\otimes D)\Big\})& \nonumber\\
&+\sigma_\Delta(A\otimes \Big\{
p(a,c)p(b,c)\N(\P(b,a),
%\Big(\!\int_{-T}^0d\lambda[b\ _\lambda\ a]\Big)\!,
c\otimes D)) 
- c\otimes\N(\P(b,a),
%\!\Big(\!\int_{-T}^0d\lambda[b\ _\lambda\ a]\Big)\!,
D)
\Big\})\ .& \nonumber
\end{eqnarray}
It follows by Corollary \ref{basic} that there exists $\Delta^\prime<\Delta$ such that
\begin{eqnarray*}
&A\otimes a\otimes\N(\P(c,b),D)
 - p(a,b)p(a,c)A\otimes\N(\P(c,b),a\otimes D)& \\
\vphantom{\Big(}
 &\equiv A\otimes\N(\P(a,\P(c,b)),D)\quad  \mod\M_{\Delta^\prime}(R)\ ,& \\
&A\otimes b\otimes\N(\P(c,a),D)
 - p(a,b)p(b,c)A\otimes\N(\P(c,a),b\otimes D)& \\
\vphantom{\Big(}
 &\equiv A\otimes\N(\P(b,\P(c,a)),D)\quad \mod\M_{\Delta^\prime}(R)\ ,& \\
&A\otimes c\otimes\N(\P(b,a),D)
 - p(b,c)p(a,c)A\otimes\N(\P(b,a),c\otimes D)& \\
\vphantom{\Big(}
 &\equiv A\otimes\N(\P(c,\P(b,a)),D)\quad \mod\M_{\Delta^\prime}(R)\ .&
\end{eqnarray*}
We can thus use the inductive assumption to rewrite (\ref{jacobson8}) as
\begin{eqnarray}
&\sigma_\Delta\Big(
p(b,c)A\otimes\N(\P(b,\P(c,a)),D)
-p(a,b)p(a,c)A\otimes\N(\P(a,\P(c,b)),D) &\nonumber\\
&-A\otimes\N(\P(c,\P(b,a)),D)
\Big) .& \label{jacobson9}
\end{eqnarray}
By Corollary \ref{lie}, the argument of $\sigma_\Delta$ in (\ref{jacobson9})
lies in $\M_{\Delta^\prime}(R)$ for some $\Delta^\prime<\Delta$ and therefore 
(\ref{jacobson9}) is zero by the inductive assumption.
This concludes the proof of the lemma.
\end{proof}

% conclusive lemma

\begin{lemma}
If $\sigma$ is as in Lemma \ref{l2}, the induced map
$$
\hat{\sigma}\ :\ U(R)=\T(R)/\M(R)\ \widetilde{\longrightarrow}\ \tU
$$
is an isomorphism of vector spaces (namely $\M(R)=\ker\sigma$).
\end{lemma}
\begin{proof}
By definition the map 
$\hat{\sigma}:\ U(R)\rightarrow\tU$ is surjective.
On the other hand, we have a natural map
$\hat{\pi}:\ \tU\rightarrow U(R)$
which maps every basis element $A=a_{i_1}\otimes\dots\otimes a_{i_n}\in\tB$
to the corresponding 
$\hat{\pi}(A)=\ :a_{i_1}\dots a_{i_n}:\ \in\B\subset U(R)$.
By Lemma \ref{l1} this map is also surjective.
The composition map
$$
\tU\ \stackrel{\hat{\pi}}{\longrightarrow}\ U(R)\
 \stackrel{\hat{\sigma}}{\longrightarrow}\ \tU
$$
is the identity map (by definition of $\hat{\pi}$ and $\hat{\sigma}$.)
This of course implies that both $\hat{\pi}$ and $\hat{\sigma}$
are isomorphisms of vector spaces.
\end{proof}
The last lemma implies that $\B$ is a basis of the space $U(R)$,
thus concluding the proof of Theorem \ref{main}.

%%%%%%%% SECTION 6 %%%%%%%%%%%%%%%%%%%%%%%%%%%%%

\section{Pre--graded freely generated vertex algebras and corresponding 
non--linear Lie conformal algebras}
\setcounter{equation}{0}

%%% preliminary constructions

Let $V$ be a vertex algebra strongly generated by a free $\C[T]$--module
$R=\C[T]\otimes\bar{V}$. Let
$$
\bar{\A}\ =\ \{\bar{a}_{\bi},\ \bi\in\bar{\I}\}\ ,
$$
be an ordered basis of $\bar{V}$, and
extend it to an ordered basis of $R$,
$$
\A\ =\ \{a_i\ ,\ \ i\in\I\}\ ,
$$
where $\I=\bar{\I}\times\Z_+$ and, if $i=(\bi,n)$, then $a_i=T^n\bar{a}_{\bi}$.
% We consider the indices $i=(\bi,k)$ ordered lexicographically.
The corresponding collection of ordered monomials of $V$ (see Definition \ref{new_free})
is denoted by $\B$. 
Recall that $V$ is said to be freely generated by $R$ if $\B$ is a basis of $V$.
We will denote by $\pi$ the natural quotient map 
$\pi\ :\ \T(R)\ \longrightarrow\ V$,
given by 
$\pi(a\otimes b\otimes\cdots\otimes c)\ =\ :ab\cdots c:$.
We will assume that the generating space $\bar{V}$ is graded by $\Gamma\backslash\{0\}$,
and that the basis $\bar{\A}$ of $\bar{V}$ is compatible 
with the $\Gamma\backslash\{0\}$--gradation.
The $\Gamma$--gradation can be extended to $R=\C[T]\otimes\bar{V}\subset V$
by saying that $T$ has zero degree,
and to the whole tensor algebra $\T(R)$ by additivity of the tensor product
(as defined in Section 2).
The corresponding $\Gamma$--filtration of $\T(R)$ naturally induces a $\Gamma$--filtration
on the vertex algebra $V$:
\begin{equation}\label{filtr}
V_\Delta\ =\ \pi(\T_\Delta(R))\ ,\quad \Delta\in\Gamma\ .
\end{equation}

%%% def of pre-graded va

\begin{definition}\label{qgrad}
The vertex algebra $V$,
strongly generated by a free $\C[T]$--module $\C[T]\otimes\bar{V}$, 
where $\bar{V}=\oplus_{\Delta\in\Gamma\backslash\{0\}}\bar{V}[\Delta]$,
is said to be {\itshape pre--graded} by $\Gamma$ if
the $\lambda$--bracket satisfies the {\itshape grading condition}
$$
[\bar{V}[\Delta_1]\ _\lambda\ \bar{V}[\Delta_2]]\subset\C[\lambda]\otimes V_{\Delta^\prime}\ ,
\quad \text{ for some }\ \Delta^\prime<\Delta\ .
$$
\end{definition}

%%% example: conf weight

\begin{example}\label{cw}
The most important examples of pre--graded vertex algebras are provided
by graded vertex algebras.
Recall that a vertex algebra $V$ is called graded if there exists a 
diagonalizable operator $L_0$ on $V$ with discrete non--negative spectrum $\Gamma$
and 0--th eigenspace $\C\vac$,
such that for all $a\in V$,
$$
[L_0,Y(a,z)]\ =\ z\partial_z Y(a,z)+Y(L_0a,z)
$$
The eigenvalue $\cw(a)$ of an eigenvector $a$ of $L_0$ is called 
its {\itshape conformal weight}.
Let $V=\oplus_{\cw\in\Gamma}V(\cw)$ be the eigenspace decomposition.
Recall that conformal weights satisfy the following rules (see e.g. \cite{kac1})
\begin{equation}\label{weights}
\cw(Ta)\ =\ \cw(a)+1\ \ ,\ \ \ \cw(a_{(n)}b)\ =\ \cw(a)+\cw(b)-n-1\ .
\end{equation}
Assume now that $V$ is strongly generated by a free $\C[T]$--submodule
$R=\C[T]\otimes\bar{V}$, where $\bar{V}$ is invariant under the action of $L_0$.
Consider the decomposition of $\bar{V}$ in a direct sum of eigenspaces of $L_0$:
$\bar{V}=\oplus_{\cw\in\Gamma\backslash\{0\}}\bar{V}[\Delta]$, 
where $\bar{V}[\Delta]=\bar{V}\cap V(\Delta)$.
This is a $\Gamma\backslash\{0\}$--gradation of $\bar{V}$, which induces, as explained above,
a $\Gamma$--gradation on $\T(R)$ and hence a $\Gamma$--filtration,
$V_\Delta,\ \Delta\in\Gamma$, on the vertex algebra $V$.
Notice that such $\Gamma$--filtration is not induced by the gradation given by the
conformal weights. Indeed the filtered space $V_\Delta$ are preserved by $T$,
while, by (\ref{weights}), $T$ increases the conformal weight by 1.
On the other hand, we have
% we can say that the filtration $V_\Delta$ is bounded by the filtration
% given by conformal weights, namely
\begin{equation}\label{weig2}
\bigoplus_{\cw\leq\Delta}V(\cw)\ \subset\ V_\Delta\ .
\end{equation}
The above $\Gamma\backslash\{0\}$--gradation of $\bar{V}$ is a
pre--gradation of $V$. Indeed (\ref{weights}) and (\ref{weig2}), imply
$$
[\bar{V}[\Delta_1]\ _\lambda\ \bar{V}[\Delta_2]]
\ \subset\ \C[\lambda]\otimes \bigoplus_{\cw\leq\Delta_1+\Delta_2-1}V(\cw)
\ \subset\  \C[\lambda]\otimes V_{\Delta_1+\Delta_2-1}\ ,
$$
so that the grading condition in Definition \ref{qgrad} holds.

\end{example}

%%% converse theorem

\begin{remark}
Notice that if $R=\C[T]\otimes \bar{V}$ is a non--linear Lie conformal algebra graded 
by $\Gamma\backslash\{0\}$,
then by Theorem \ref{main} the universal enveloping vertex algebra $U(R)$ is a pre--graded 
freely generated vertex algebra.
\end{remark}
We want to prove a converse statement to Theorem \ref{main}.
\begin{theorem}\label{converse}
Let $V$ be a pre--graded vertex algebra freely generated by a free
$\C[T]$--submodule $R=\C[T]\otimes\bar{V}$.
Then $R$ has a structure of non--linear Lie conformal algebra
$$
\P_\lambda\ :\ \ R\otimes R\ \longrightarrow\ \C[\lambda]\otimes\T(R)\ ,
$$
compatible with the Lie conformal algebra structure of $V$, in the sense that
$$
\pi(\P_\lambda(a,b))\ =\ [a\ _\lambda\ b]\ ,\ \ \forall a,b\in R\ ,
$$
so that $V$ is canonically isomorphic to the universal enveloping vertex algebra $U(R)$.
\end{theorem}
The proof will be achieved as a result of 8 lemmas.

% lemma 1

\begin{lemma}\label{fin1}
The vertex algebra structure on $V$ satisfies the following grading conditions:
\begin{eqnarray*}
\vphantom{\Big(}
:V_{\Delta_1}V_{\Delta_2}: &\subset& V_{\Delta_1+\Delta_2}\ ,\\
\vphantom{\Big(}
[V_{\Delta_1}\ _\lambda\ V_{\Delta_2}] &\subset& \C[\lambda]\otimes V_{\Delta^\prime}
\ \text{ for some } \Delta^\prime<\Delta_1+\Delta_2\ .
\end{eqnarray*}
\end{lemma}
\begin{proof}
By definition $V_\Delta$ is spanned by elements $\pi(A)$, where $A\in\T_\Delta(R)$.
We thus want to prove, by induction on $\Delta=\Delta_1+\Delta_2$, that,
for $A\in\T_{\Delta_1}(R)$ and $B\in\T_{\Delta_2}(R)$, we have
\begin{eqnarray}
\vphantom{\Big(}
:\pi(A)\pi(B): &\in& V_{\Delta}\ ,\label{pr1_fin1}\\
\vphantom{\Big(}
[\pi(A)\ _\lambda\ \pi(B)] &\in& \C[\lambda]\otimes V_{\Delta^\prime}
\ \text{ for some } \Delta^\prime<\Delta\ .\label{pr2_fin1}
\end{eqnarray}
If $A\in\C$ or $B\in\C$, both conditions are obvious.
Suppose then $A,B\in R\oplus R^{\otimes2}\oplus\cdots$.
%1
If $A\in R$, we have $:\pi(A)\pi(B):\ =\ \pi(A\otimes B)$,
so that condition (\ref{pr1_fin1}) follows immediately by the definition 
of the $\Gamma$--filtration on $V$.
Let then $A=a\otimes A^\prime$, with $a\in R$ and $A^\prime,B\in R\oplus R^{\otimes2}\oplus\cdots$.
Notice that
$\Delta(a)+\Delta(A^\prime)=\Delta(A)$ and $\Delta(a),\Delta(A^\prime)>0$.
By quasi--associativity of the normally ordered product we then have
\begin{eqnarray*}
:\pi(A)\pi(B): &=& :a(:\pi(A^\prime)\pi(B):):
+:\Big(\int_0^Td\lambda\ a\Big) [\pi(A^\prime)\ _\lambda\ \pi(B)]: \\
&+& p(a,A^\prime):\Big(\int_0^Td\lambda\ \pi(A^\prime)\Big) [a\ _\lambda\ \pi(B)]:\ , 
\end{eqnarray*}
and each term in the right hand side belongs to $V_\Delta$ by the inductive assumption.
%2
We are left to prove condition (\ref{pr2_fin1}).
If $A,B\in R$, (\ref{pr2_fin1}) holds by sesquilinearity and
by the assumption of grading condition on $V$.
Suppose now $A=a\in R$ and $B=b\otimes B^\prime$, 
with $b\in R$ and $B^\prime\in R\oplus R^{\otimes2}\oplus\cdots$.
In this case we can use the left Wick formula to get
\begin{eqnarray*}
[\pi(A) \ _\lambda \ \pi(B)] &=& 
:[a \ _\lambda \  b]\pi(B^\prime):+p(a,b):b[a \ _\lambda \  \pi(B^\prime)]: \\
&+& \int_0^\lambda d\mu [[a \ _\lambda \  b] \ _\mu \  \pi(B^\prime)] \ ,
\end{eqnarray*}
and each term in the right hand side belongs to $V_{\Delta^\prime}$,
for some $\Delta^\prime<\Delta$, by induction.
Finally, let $A=a\otimes A^\prime$, where $a\in R$ 
and $A^\prime,B\in R\oplus R^{\otimes2}\oplus\cdots$.
We then have by the right Wick formula
\begin{eqnarray*}
[\pi(A) \ _\lambda \ \pi(B)] &=&
:\left(e^{T\partial_\lambda}a\right)[\pi(A^\prime) \ _\lambda \  \pi(B)]:
+p(a,A^\prime):\left(e^{T\partial_\lambda}\pi(A^\prime)\right)[a \ _\lambda \  \pi(B)]: \\ 
&+& p(a,A^{\prime})\int_0^\lambda d\mu [\pi(A^\prime) \ _\mu \ [a \ _{\lambda-\mu} \  \pi(B)]] \ ,
\end{eqnarray*}
and again, by the inductive assumption, each term in the right hand side belongs to
$V_{\Delta^\prime}$ for some $\Delta^\prime<\Delta$.
\end{proof}

%%% primary gradation

As in Section 5, we denote by $\tB$ the collection of ordered monomials in $\T(R)$:
$$
\tB\ =\
\bigg\{
a_{i_1}\otimes\dots\otimes a_{i_n}
\left|\begin{array}{c}
i_1\leq\dots\leq i_n,\  n\in\Z_+\ ,\\
i_q<i_{q+1} \text{ if } p(a_{i_q})=\bar{1}
\end{array}\right\}\ \subset\ \T(R)\ ,
$$
so that the basis $\B$ of $V$ is the image of $\tB$ under the quotient map $\pi$:
$$
\B\ =\
\bigg\{
:a_{i_1}\dots a_{i_n}:
\left|\begin{array}{c}
i_1\leq\dots\leq i_n,\  n\in\Z_+\ ,\\
i_q<i_{q+1} \text{ if } p(a_{i_q})=\bar{1}
\end{array}\right\}\ \subset\ V\ .
$$
The basis $\B$ of $V$ induces an embedding $\rho:\ V\hookrightarrow\T(R)$,
defined by
$$
\rho(:a_{i_1}\dots a_{i_n}:)\ =\ a_{i_1}\otimes\dots\otimes a_{i_n}\ ,
\quad\forall :a_{i_1}\dots a_{i_n}:\in\B\ .
$$
Moreover, the basis $\B$ induces a {\itshape primary $\Gamma$--gradation} on $V$,
\begin{equation}\label{prim_grad}
V\ =\ \bigoplus_{\Delta\in\Gamma}V[\Delta]^{(1)}\ ,
\end{equation}
defined by assigning to every basis element $:a_{i_1}\dots a_{i_n}:\ \in\B$ degree
$$
\Delta(:a_{i_1}\dots a_{i_n}:)\ =\ \Delta(a_{i_1})+\dots+\Delta(a_{i_n})\ .
$$

%%% lemma 2

\begin{lemma}\label{fin2}
The $\Gamma$--filtration (\ref{filtr}) of $V$ is induced 
by the primary $\Gamma$--gradation (\ref{prim_grad}).
\end{lemma}
\begin{proof}
We want to prove that for every $\Delta\in\Gamma$ 
$$
V_\Delta=\bigoplus_{\Delta^\prime\leq\Delta}V[\Delta^\prime]^{(1)}\ .
$$
By definition of the $\Gamma$--filtration of $V$
$$
V_\Delta=\pi(\T_\Delta(R))\
=\ \Span_\C\Big\{:a_{j_1}\dots a_{j_n}:\ \Big|\ \Delta(a_{j_1})+\dots+\Delta(a_{j_n})\leq\Delta\Big\}\ ,
$$
while, by definition of the primary $\Gamma$--gradation of $V$, we have
$$
\bigoplus_{\Delta^\prime\leq\Delta}V[\Delta^\prime]^{(1)}\ =\
\Span_\C\left\{\begin{array}{c}
\\
:a_{j_1}\dots a_{j_n}:\\
\end{array}\right.
\left|\begin{array}{c}
\Delta(a_{j_1})+\dots+\Delta(a_{j_n})\leq\Delta \\
i_1\leq\cdots\leq i_n \\
i_q<i_{q+1}\ \text{ if } p(a_{i_q})=\bar{1} 
\end{array}\right\}\ .
$$
Therefore we obviously have the inclusion
$\bigoplus_{\Delta^\prime\leq\Delta}V[\Delta^\prime]^{(1)}\subset V_\Delta$.
We are left to show that every (non necessarily ordered) monomial
$A=:a_{j_1}\dots a_{j_n}:$
% belongs to $\bigoplus_{\Delta^\prime\leq\Delta}V[\Delta^\prime]^{(1)}$,
% if $\Delta(a_{j_1})+\dots+\Delta(a_{j_n})\leq\Delta$.
% In other words, we need to prove that $A$ 
can be written as linear combination
of ordered monomials of primary degree $\Delta^\prime\leq\Delta$.
We will prove this statement by induction on $(\Delta,d)$,
where $d$ is the number of inversions of $A$:
$$
d=\#\left\{(p,q)\ \ \left|\ \ \begin{array}{c}
1\leq p<q\leq n \text{ and either } j_p>j_q \\
\text{ or } j_p=j_q \text{ and } p(a_{j_p})=\bar{1}
\end{array}\right.\right\}\ .
$$
If $d=0$, then $A\in V[\Delta]^{(1)}$ and the statement is trivial.
Suppose then $d\geq1$, and let $(j_q,j_{q+1})$ be an inversions of $A$.
We consider the case $j_q>j_{q+1}$. The case $j_q=j_{q+1}$ and $p(a_{j_q})=\bar{1}$
is similar.
By skew--symmetry of the normally ordered product, we have
\begin{eqnarray}\label{pr1_fin2}
A &=& p(a_{j_q},a_{j_{q+1}}):a_{j_1}\cdots a_{j_{q+1}}a_{j_q}\cdots a_{j_n}: \\
&+& a_{j_1}\cdots a_{j_{q-1}}
\Big(\int_{-T}^0d\lambda\ [a_{j_q}\ _\lambda\ a_{j_{q+1}}]\Big)
a_{j_{q+2}}\cdots a_{j_n}:\ .\nonumber
\end{eqnarray}
The first term in the right hand side of (\ref{pr1_fin2}) is a monomial
of $V_\Delta$ with $d-1$ inversions.
Moreover, by Lemma \ref{fin1}, the second term in the right hand side of (\ref{pr1_fin2})
belongs to $V_{\Delta^\prime}$ for some $\Delta^\prime<\Delta$.
We can thus use the inductive assumption to conclude, as we wanted, that
$A\in\bigoplus_{\Delta^\prime\leq\Delta}V[\Delta^\prime]^{(1)}$.
\end{proof}

%%% new basis

The embedding $\rho:\ V\hookrightarrow\T(R)$ introduced above 
does not commute with the action of $T$.
The main goal in the following will be to replace $\rho$ with another embedding 
$\rho_T:\ V\hookrightarrow\T(R)$ which commutes with the action of $T$.

We introduce the following collection of elements of $\T(R)$
$$
\tB_T\ =\
\bigg\{
T^{k_1}(\bar{a}_{\bi_1}\otimes a_{i_2}\otimes\dots\otimes a_{i_n})
\left|\begin{array}{c}
i_1=(\bi_1,k_1)\leq i_2\leq\dots\leq i_n \\
i_q<i_{q+1} \text{ if } p(a_{i_q})=\bar{1}
\end{array}\right\}\ \subset\ \T(R)\ ,
$$
and we denote by $\B_T$ the corresponding image via the quotient map $\pi$,
namely
$$
\B_T\ =\
\bigg\{
T^{k_1}:\bar{a}_{\bi_1}a_{i_2}\dots a_{i_n}:
\left|\begin{array}{c}
i_1=(\bi_1,k_1)\leq i_2\leq\dots\leq i_n\\
i_q<i_{q+1} \text{ if } p(a_{i_q})=\bar{1}
\end{array}\right\}\ \subset\ V\ .
$$

%%% lemma 3

\begin{lemma}\label{fin3}
Consider the elements $:a_{i_1}\dots a_{i_n}:\ \in\B$ 
and $T^{k_1}:\bar{a}_{\bi_1}a_{i_2}\dots a_{i_n}:\ \in\B_T$ ordered lexicographically by the indices
$(\Delta,i_1=(\bi_1,k_1),i_2=(\bi_2,k_2),\dots,i_n=(\bi_n,k_n))$,
where $\Delta=\Delta(a_{i_1})+\dots+\Delta(a_{i_n})\in\Gamma$.
\smallskip

\noindent (a) Every element $T^{k_1}:\bar{a}_{\bi_1}a_{i_2}\dots a_{i_n}:\ \in\B_T$
can be written as
$$
T^{k_1}:\bar{a}_{\bi_1}a_{i_2}\dots a_{i_n}:
\ =\ :a_{i_1}\dots a_{i_n}:+M\ ,
$$
where $M$ is linear combination of elements of $\B$ smaller than $:a_{i_1}\dots a_{i_n}:$.
% (according to the above ordering of $\B$).
\smallskip

\noindent (b) Every element $:a_{i_1}\dots a_{i_n}:\ \in\B$
can be written as
$$
:a_{i_1}\dots a_{i_n}:
\ =\ T^{k_1}:\bar{a}_{\bi_1}a_{i_2}\dots a_{i_n}:+N\ ,
$$
where $N$ is linear combination of elements of $\B_T$ smaller than 
$T^{k_1}:\bar{a}_{\bi_1}a_{i_2}\dots a_{i_n}:$.
\end{lemma}
\begin{proof}
% a
(a) Since  $T$ is a derivation of the normally ordered product, we have
\begin{eqnarray*}
\vphantom{\Big(}
&&\qquad\qquad
 T^{k_1}:\bar{a}_{\bi_1}a_{i_2}\dots a_{i_n}:\
=\ :a_{i_1}\dots a_{i_n}: \\
&& + \sum_{\begin{array}{c}
l_1\geq1,l_2,\dots,l_n\geq0 \\
l_1+\cdots+l_n=k_1
\end{array}}
\left(\begin{array}{c}
k_1 \\
l_1,\cdots,l_n
\end{array}\right)
:a_{(\bi_1,l_1)}a_{(\bi_2,k_2+l_2)}\cdots a_{(\bi_n,k_n+l_n)}:\ . 
\end{eqnarray*}
We just need to prove that every monomial 
$:a_{(\bi_1,l_1)}a_{(\bi_2,k_2+l_2)}\cdots a_{(\bi_n,k_n+l_n)}:$
is linear combination of ordered monomials smaller than 
$:a_{i_1}\cdots a_{i_n}:$.
In general the monomial
$:a_{(\bi_1,l_1)}a_{(\bi_2,k_2+l_2)}\cdots a_{(\bi_n,k_n+l_n)}:$ is not ordered.
But thanks to skew--symmetry of the normally ordered product and to Lemma \ref{fin1},
we can rewrite it (up to a sign) as sum of an ordered monomial $:a_{\sigma_1}\cdots a_{\sigma_n}:$
with the same (reordered) indices, and an element $R$ of $V_{\Delta^\prime}$,
with $\Delta^\prime<\Delta$.
Notice that the reordered monomial $:a_{\sigma_1}\cdots a_{\sigma_n}:$ is smaller
than $:a_{i_1}\cdots a_{i_n}:$, since $\Delta(a_{\sigma_1})+\cdots+\Delta(a_{\sigma_n})=\Delta$
and $\sigma_1=(\bi_1,l_1)<i_1$.
Moreover, by Lemma \ref{fin2}, $R$ is linear combination of ordered monomials
with primary degree $\Delta^\prime<\Delta$, hence smaller than $:a_{i_1}\cdots a_{i_n}:$.
This proves the first part of the lemma.
\smallskip %b

\noindent (b) As before, we can decompose
\begin{eqnarray*}
\vphantom{\Big(}
&&\qquad\qquad
 :a_{i_1}\dots a_{i_n}: \ =\ T^{k_1}:\bar{a}_{\bi_1}a_{i_2}\dots a_{i_n}: \\
&& - \sum_{\begin{array}{c}
l_1\geq1,l_2,\dots,l_n\geq0 \\
l_1+\cdots+l_n=k_1
\end{array}}
\left(\begin{array}{c}
k_1 \\
l_1,\cdots,l_n
\end{array}\right)
:a_{(\bi_1,l_1)}a_{(\bi_2,k_2+l_2)}\cdots a_{(\bi_n,k_n+l_n)}:\ . 
\end{eqnarray*}
and we want to prove that every monomial 
$:a_{(\bi_1,l_1)}a_{(\bi_2,k_2+l_2)}\cdots a_{(\bi_n,k_n+l_n)}:$
is linear combination of elements of $\B_T$ smaller 
than $T^{k_1}:\bar{a}_{\bi_1}a_{i_2}\dots a_{i_n}:$.
By the above argument, we can decompose
$$
:a_{(\bi_1,l_1)}a_{(\bi_2,k_2+l_2)}\cdots a_{(\bi_n,k_n+l_n)}:
\ =\ :a_{\sigma_1}a_{\sigma_2}\cdots a_{\sigma_n}:+R\ ,
$$
where $:a_{\sigma_1}a_{\sigma_2}\cdots a_{\sigma_n}:\in\B$
is the ordered monomial obtained by reordering the indices of 
$:a_{(\bi_1,l_1)}a_{(\bi_2,k_2+l_2)}\cdots a_{(\bi_n,k_n+l_n)}:$,
and $R\in V_{\Delta^\prime}$, for some $\Delta^\prime<\Delta$.
Since $(\bi_1,l_1)<i_1\leq(i_q,k_q+l_q)\ ,\ \forall q=2,\dots,n$,
it follows by induction on $(\Delta,i_1,\dots,i_n)$ that $:a_{\sigma_1}\cdots a_{\sigma_n}:$
is linear combination of elements of $\B_T$ smaller than 
$T^{k_1}:\bar{a}_{\bi_1}a_{i_2}\dots a_{i_n}: $.
By Lemma \ref{fin2} $R$ is a linear combination of ordered monomials
$:a_{\tau_1}\cdots a_{\tau_p}:\ \in\B$, with $\Delta(a_{\tau_1})+\cdots+\Delta(a_{\tau_p})<\Delta$.
We thus conclude, by induction on $(\Delta,i_1,\dots,i_n)$, that each such
ordered monomial is linear combination of elements of $\B_T$ smaller than
$T^{k_1}:\bar{a}_{\bi_1}a_{i_2}\dots a_{i_n}: $.
\end{proof}

%%% lemma 4

\begin{lemma}\label{fin4}
$\B_T$ is a basis of $V$.
\end{lemma}
\begin{proof}
By Lemma \ref{fin3}(b), $\B_T$ spans $V$.
Suppose by contradiction that there is a relation of linear dependence among elements of $\B_T$,
and let $T^{k_1}:\bar{a}_{\bi_1}a_{i_2}\dots a_{i_n}: $ be the largest element of $\B_T$
(with respect to the ordering defined in Lemma \ref{fin3}) with a non zero coefficient $c$.
By Lemma \ref{fin3}(a), we can rewrite such relation as a relation of linear dependence among
elements of $\B$, which is not identically zero since the coefficient of 
$:a_{(\bi_1,k_1)}a_{i_2}\dots a_{i_n}:$ is $c\neq0$.
This clearly contradicts the fact that $\B$ is a basis of $V$.
\end{proof}

%%% secondary gradation

The basis $\B_T$ of $V$ induces a new embedding $\rho_T:\ V\hookrightarrow\T(R)$,
defined by
$$
\rho_T(T^{k_1}:\bar{a}_{\bi_1}a_{i_2}\dots a_{i_n}:)\ 
=\ T^{k_1}(\bar{a}_{\bi_1}\otimes a_{i_2}\otimes\dots\otimes a_{i_n})\ ,
$$
for every basis element
$T^{k_1}:\bar{a}_{\bi_1}a_{i_2}\dots a_{i_n}:\ \in\B_T$.
Notice that $\rho_T$ commutes with the action of $T$.
Moreover, $\B_T$ induces a {\itshape secondary $\Gamma$--gradation} on $V$,
\begin{equation}\label{sec_grad}
V\ =\ \bigoplus_{\Delta\in\Gamma}V[\Delta]^{(2)}\ ,
\end{equation}
defined by assigning to every basis element 
$T^{k_1}:\bar{a}_{\bi_1}a_{i_2}\dots a_{i_n}:\ \in\B_T$
degree
$$
\Delta(T^{k_1}:\bar{a}_{\bi_1}a_{i_2}\dots a_{i_n}:)
\ =\ \Delta(a_{i_1})+\dots+\Delta(a_{i_n})\ .
$$

%%% lemma 5

\begin{lemma}\label{fin5}
(a) The $\Gamma$--filtration (\ref{filtr}) of $V$ is induced 
by the secondary $\Gamma$--gradation (\ref{sec_grad}).
\smallskip

\noindent (b) The embedding $\rho_T$ preserves the filtration: 
$\rho_T(V_\Delta)\subset\T_\Delta(R)$.
\end{lemma}
\begin{proof}
% Thanks to Lemma \ref{fin2}, we just need to show that the primary gradation (\ref{prim_grad})
% and the secondary gradation (\ref{sec_grad}) induce the same filtration, namely
% $$
% \bigoplus_{\Delta^\prime\leq\Delta}V[\Delta^\prime]^{(1)}
% \ =\ \bigoplus_{\Delta^\prime\leq\Delta}V[\Delta^\prime]^{(2)}\ .
% $$
% But this
The first part of the lemma is an obvious corollary of Lemma \ref{fin3}.
The second part follows by the first part and the obvious inclusion
$\rho_T(V[\Delta]^{(2)})\subset\T(R)[\Delta],\ \forall\Delta\in\Gamma$.
\end{proof}

%%% lemma 6

\begin{lemma}\label{fin6}
(a) $R$ is a non--linear skew--symmetric conformal algebra with $\lambda$--bracket
$$
\P_{\lambda}\ :\ \ R\otimes R\ \longrightarrow\ \C[\lambda]\otimes\T(R)\ ,
$$
given by
$$
\P_\lambda(a,b)\ =\ \rho_T([a\ _\lambda\ b])\ ,\quad
\forall a,b\in R.
$$
(b) Such $\lambda$--bracket is compatible with $[\ _\lambda\ ]$, namely
\begin{eqnarray*}
\pi(\P_\lambda(A,B)) &=& [\pi(A)\ _\lambda\ \pi(B)]\ ,\\
\pi(\N(A,B)) &=& :\pi(A)\pi(B):\ ,\qquad \forall A,B\in\T(R)\ .
\end{eqnarray*}
Here $\P_\lambda$ and $\N$ are defined on $\T(R)$
thanks to Lemma \ref{extension}.
\end{lemma}
\begin{proof}
(a) Since $\rho_T$ commutes with the action of $T$,
it immediately follows that $\P_\lambda$ satisfies sesquilinearity and skew--symmetry.
The grading condition for $\P_\lambda$
follows by the assumptions on $V$ and Lemma \ref{fin5}(b).
Indeed
$$
\P_\lambda(R[\Delta_1],R[\Delta_2])
\ =\ \rho_T([R[\Delta_1]\ _\lambda\ R[\Delta_2]])
\ \subset\ \C[\lambda]\otimes\rho_T(V_\Delta)
\ \subset\ \C[\lambda]\otimes\T_\Delta(R)\ ,
$$
for some $\Delta<\Delta_1+\Delta_2$.
% This proves the first part of the lemma.
\smallskip

\noindent (b) Notice that, by definition of $\pi$ and $\rho_T$,
we have $\pi\circ\rho_T=\id_V$.
Hence $\pi(\P_\lambda(a,b)) = [a\ _\lambda\ b],\ \forall a,b\in R$.
Part (b) then follows by the definition of $\P_\lambda$ and $\N$ on $\T(R)$,
and by an easy induction argument.
\end{proof}

%%% lemma 7

Notice that in the proof of Lemma \ref{l1} we did not use the assumption that $R$ 
was a non--linear Lie conformal algebra.
As immediate consequence we get the following
\begin{lemma}\label{fin7}
Every element $A\in\T_\Delta(R)$
decomposes as 
$$
A=B+M\ ,
$$
with $B\in\Span_\C\tilde{\B}_\Delta$ and $M\in\M_\Delta(R)$
(we are using the notation introduced in Section 5).
\end{lemma}

%%% lemma 8

To conclude the proof of Theorem \ref{converse} we are left to prove
% show that $\P_\lambda$
% defines on $R$ the structure of non--linear Lie conformal algebra, namely it
% satisfies the Jacobi identity.
% This is stated in the following
\begin{lemma}\label{fin8}
The $\lambda$--bracket $\P_\lambda:\ R\otimes R\rightarrow\C[\lambda]\otimes\T(R)$
satisfies the Jacobi identity (\ref{J}).
\end{lemma}
\begin{proof}
We need to show that for elements $a,b,c$ of $R$ we have
\begin{eqnarray*}
\vphantom{\Big(}
\j(a,b,c;\lambda,\mu)
&=& \P_\lambda(a,\P_\lambda(b,c))-p(a,b)\P_\mu(b,\P_\lambda(a,c)) \\
\vphantom{\Big(}
&-&\P_{\lambda+\mu}(\P_\lambda(a,b),c)\ \in\ \M_{\Delta}(R)\ .
\end{eqnarray*}
for some $\Delta\in\Gamma$ such that $\Delta<\Delta(a)+\Delta(b)+\Delta(c)$.
By the grading condition (\ref{8}) on $\P_\lambda$, we have
$$
\j(a,b,c;\lambda,\mu)\in\T_\Delta(R)\ ,
$$
for some $\Delta<\Delta(a)+\Delta(b)+\Delta(c)$.
Therefore, by Lemma \ref{fin7}, we have
\begin{equation}\label{f1}
\j(a,b,c;\lambda,\mu)=\j_B+\j_M\ ,
\end{equation}
where $\j_B\in\Span_\C\tilde{\B}_\Delta$ and $\j_M\in\M_\Delta(R)$.
By Lemma \ref{fin6}(b) $\P_\lambda$ is compatible with $[\ _\lambda\ ]$.
Since $V$ is a vertex algebra, we have
\begin{eqnarray}\label{f2}
\vphantom{\Big(}
\pi(\j(a,b,c;\lambda,\mu))
&=& [a\ _\lambda\ [b\ _\mu\ c]]-p(a,b) [b\ _\mu\ [a\ _\lambda\ c]] \nonumber\\
\vphantom{\Big(}
&-& [[a\ _\lambda\ b]\ _{\lambda+\mu}\ c]\ =\ 0\ .
\end{eqnarray}
Moreover, since $\M(R)\subset\Ker\pi$, we also have
\begin{equation}\label{f3}
\pi(\j_M)\ =\ 0\ .
\end{equation}
We thus get, from (\ref{f1}),(\ref{f2}) and (\ref{f3}), that
$\pi(\j_B)\ =\ 0$.
On the other hand, since $V$ is freely generated over $\bar{V}$, we have that
$\pi:\ \Span_\C\tB\rightarrow V$ is an isomorphism of vector spaces.
We thus conclude, as we wanted, that $\j_B=0$, hence
$\j(a,b,c;\lambda,\mu)\in\M_\Delta(R)$.
\end{proof}

% conclusive remark

\begin{remark}
Let $V$ be any simple graded vertex algebra strongly generated 
by a $\C[T]$--submodule $R$.
Arguments similar to the ones used in the proof of Theorem \ref{converse} show that
the vertex algebra structure of $V$ induces on $R$ a structure of
a non--linear conformal algebra.
We have to consider separately two cases.
\smallskip

\noindent 1.\ If $R$ is a non--linear Lie conformal algebra, then it is not hard to show that
$V$ is the simple graded quotient of the universal enveloping vertex algebra $U(R)$
by a unique maximal "irregular" ideal.
In this case we say that the vertex algebra $V$ is {\itshape non--degenerate}.
We thus conclude that non--degenerate vertex algebras are classified
by non--linear Lie conformal algebras.
Their classification will be studied in a subsequent paper \cite{non_lin_2}.

\smallskip

\noindent 2.\ If on the contrary $R$ is a non--linear conformal algebra, but 
Jacobi identity (\ref{J})
does not hold, we say that the vertex algebra $V$ is {\itshape degenerate}.
In this case one can still describe $V$ as quotient of the universal enveloping vertex algebra $U(R)$ of $R$.
The main difference is that in general $U(R)$ is not freely generated by $R$, namely 
PBW theorem fails.
For this reason, the study (and classification) of degenerate vertex algebras is more complicated
than in the non--degenerate case.
Their structure theory will be developed in subsequent work \cite{non_lin_2}.
\end{remark}

%%%%%%%%%%%%%%%%%% BIBLIOGRAFIA %%%%%%%%%%%%%%%

%\bibliography{../../../../latex/master_bibliography}
\newcommand{\noopsort}[1]{} \newcommand{\printfirst}[2]{#1}
  \newcommand{\singleletter}[1]{#1} \newcommand{\switchargs}[2]{#2#1}
  \def\cprime{$'$}

\bibliographystyle{alpha}

\end{document}